\documentclass[format=acmsmall, review=false, screen=true]{acmart}

\usepackage{array,booktabs} 
\usepackage{adjustbox}
\usepackage{rotating}
\usepackage[ruled]{algorithm2e} 

\SetAlFnt{\small}
\SetAlCapFnt{\small}
\SetAlCapNameFnt{\small}
\SetAlCapHSkip{0pt}
\IncMargin{-\parindent}
\usepackage{multirow}
\usepackage{caption}
\usepackage{subcaption}

\acmJournal{TOCHI}

\setcopyright{acmlicensed}

\makeatletter
\def\@copyrightspace{\relax}
\makeatother
\begin{document}
\title{New tab page recommendations cause a strong suppression of exploratory web browsing behaviors}


\author{Homanga Bharadhwaj}
\authornote{This is the corresponding author}
\orcid{1234-5678-9012-3456}
\affiliation{%
 \institution{IIT Kanpur}
 \city{Kanpur}
 \state{U.P.}
 \country{India}}
\email{homangab@cse.iitk.ac.in}

\author{Nisheeth Srivastava}
\affiliation{%
 \institution{IIT Kanpur}
 \city{Kanpur}
 \state{U.P.}
 \country{India}}
\email{nsrivast@cse.iitk.ac.in}

\def\plainkeywords{filter bubble; personalization; exploration; habitual browsing; information diet; human memory; cognitive science}

\begin{abstract}
Through a combination of experimental and simulation results, we illustrate that passive recommendations encoded in typical computer user-interfaces (UIs) can subdue users' natural proclivity to access diverse information sources. Inspired by traditional demonstrations of a part-set cueing effect in the cognitive science literature, we performed an online experiment manipulating the operation of the `New Tab' page for consenting volunteers over a two month period. Examination of their browsing behavior reveals that typical frequency and recency-based methods for displaying websites in these displays subdues users' propensity to access infrequently visited pages compared to a situation wherein no web page icons are displayed on the new tab page. Using a carefully designed simulation study, representing user behavior as a random walk on a graph, we inferred quantitative predictions about the extent to which discovery of new sources of information may be hampered by personalized `New Tab' recommendations in typical computer UIs. We show that our results are significant at the individual level and explain the potential consequences of the observed suppression in web-exploration.
\end{abstract}



\maketitle

\section{Introduction}
The medium is the message~\cite{mcluhan1994understanding}. McLuhan's gnomic aphorism has, in the half-century since its utterance, launched countless humanistic appraisals of the manner in which the form of modern communication technologies have shaped, and continue to shape the way people think. Given the extent to which web-based technologies now saturate our lived environment, substantiating these critiques quantitatively is very important. 

For instance, it seems intuitively plausible, even likely, that the hyper-linked nature of web interfaces, in contrast with the linear nature of books, leads to greater distraction and shallower processing of information~\cite{carr2010shallows}. A survey of recent empirical research addressing this question, however, notes that even though the basic hypothesis is borne out - students reading material on screens do worse on tests than students reading material on paper, this performance differential appears to reduce quite rapidly with practice~\cite{myrberg2015screen}. While it is still too early to conclude definitively either way about the validity of this particular critique, a general point remains: access to empirical data permits broad critiques of technology's influence on cognition to themselves be substantiated. 

A similar story can be told about yet another intuitively appealing hypothesis - the filter bubble~\cite{pariser2011filter}. Here, the claim is that web-based recommendation services, by virtue of their algorithmic logic, recommend digital objects to people that they are already known to prefer, thus reducing the possibility for them to experience new content that they may or may not like. Again, the claim is easily believable, but has proved difficult to substantiate empirically. An analysis of the effect of accepting movie recommendations from the Movielens system showed that while the set of movies recommended became gradually less diverse with time, users who chose to follow the system's recommendations received more diverse recommendations than users who ignored them~\cite{nguyen2014exploring}. Recent research seeking to evaluate the existence of filter bubbles in Google News' personalization algorithm reported extremely minor differences in news articles presented to different user accounts explicitly set up to provoke differentiation~\cite{haim2018burst}. Mixed evidence for filter bubbles was obtained from an analysis conducted by Microsoft on the news reading habits of 50000 Internet Explorer users, which found that while political articles accessed via social media or search engines are more politically extreme than those accessed directly from websites, the use of these channels is also more associated with exposure to opposite political views than direct access~\cite{flaxman2016filter}.    

There is an interesting dichotomy between theoretical and empirical evaluations of the filter bubble. Whereas theoretical critiques focus heavily on how recommendations could potentially narrow the diversity of users' demand patterns for digital media~\cite{pariser2011filter}, empirical research focuses on identifying whether recommendation systems are narrowing the diversity of content providers' {\em supply} of digital objects~\cite{nguyen2014exploring, haim2018burst}. Work that explicitly tracks users' {\em consumption} patterns does appear to detect a small bubble effect~\cite{flaxman2016filter}, but the restriction of this study to studying only political diversity reduces the generalizability of its results. 

In this paper, we report results from an experimental study of an extremely general version of the filter bubble hypothesis, evaluated strictly on consumption patterns rather than supply, and without restricting the nature of content consumed. Specifically, we investigate whether the presence of frequently visited tabs on `new tab' pages in modern browsers leads to a concentration of browsing behavior into a more restricted set of websites than when such recommendation-based displays are absent. 

This hypothesis originated from a simple analogy: users' interaction with the `new tab page' display in modern browsers almost precisely reproduces the environmental conditions under which psychological test subjects experience memory inhibition in part-set cueing experiments~\cite{slamecka1968examination}. In these classic experiments, when asked to learn a set of words for future recall, people performed worse during memory testing when they were actually shown some subset of the words than when they were simply asked to remember as many words as they could. 

Analogously, in principle an Internet user browsing the web can go to any URL they remember (or can type the first few letters of). This maps on to the free recall memory task. The user's attention is drawn to a set of page icons on the new tab page during the recall task, reproducing the partial set presentation condition of part-set cueing experiments. If the lab finding translates to this natural setting, we'd expect that users' propensity to access non-displayed pages would be impaired by the mere presence of web page icons in the new tab page display.

 We conducted a controlled, longitudinal, within-subject experiment wherein we manipulated the behavior of consenting volunteers' new tab page displays without their knowledge and analyzed their browsing history at the end of the observation period to measure the degree to which our covert manipulations influenced users' behavior. We found strong statistical support for our hypothesis. Usage of new tab pages with personalized recommendations considerably reduced the number of unique websites visited by users, suggesting concentration of website browsing repertoires. Additional analyses revealed that this concentration was not uniform, and was seen most clearly precisely during browsing events most strongly associated with free memory recall - users typing URLs into address bars in the new tab page display, with recommendations visible underneath.
 
 To assess the extent to which this measured deterioration in exploratory browsing could affect the diversity of users' information foraging on the web, we also ran a simulation study, using our data to set parameters for a random walk model of a web surfer seeking information from diverse sources. We showed that, for empirically grounded values of exploratory browsing, the size of suppression effects seen in our empirical sample would correspond to reductions in the frequency with which diverse information sources are accessed, of nearly 50\% for the median user. 
 
 Thus, in this paper, we present empirical and simulation evidence to show that the mere presence of a set of web page icons on browser new tab pages reduces users' propensity to access infrequently visited web pages by nearly half. We also discuss the implications of this striking finding on the filter bubble literature, the principles of designing communication interfaces, and ethical considerations of web mediated communication.

 \section{A controlled filter bubble experiment}
 
 \begin{center}
\begin{figure}[htbp]
\includegraphics[width=0.95\textwidth]{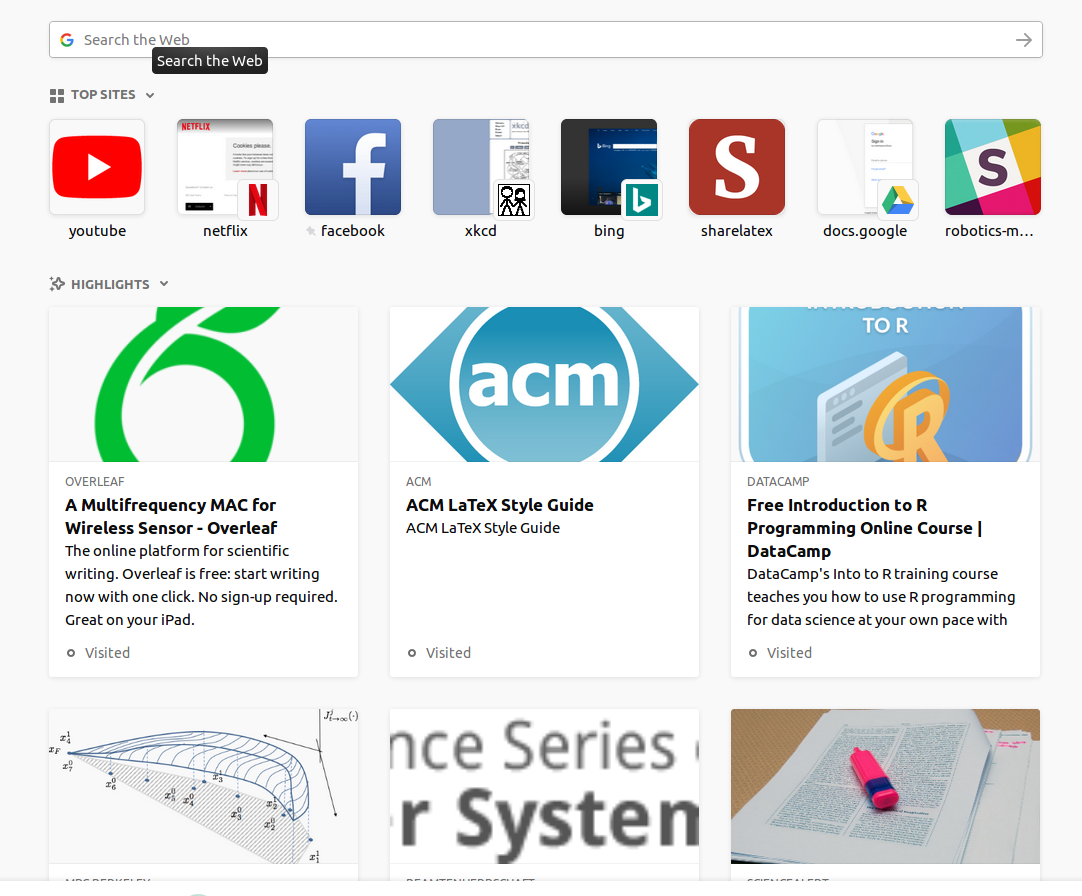}
\caption{A typical new tab page with different web page icons shown. Although the layout differs for different web-browsers and different versions of the same browser, the basic premise of displaying different web page icons persists across browsers.}
\label{fig:example}
\end{figure}
\end{center}

We anticipated large individual differences in browsing behavior across subjects, suggesting that  cohort-level differences may or may not be representative of the median user's experience. Therefore, we focused on designing a within-subject experiment, with experimental manipulations that each participant in the experiment would experience in different time blocks.




\subsection{Experiment design} 
\begin{center}
\begin{figure}[htbp]
\includegraphics[width=0.95\columnwidth]{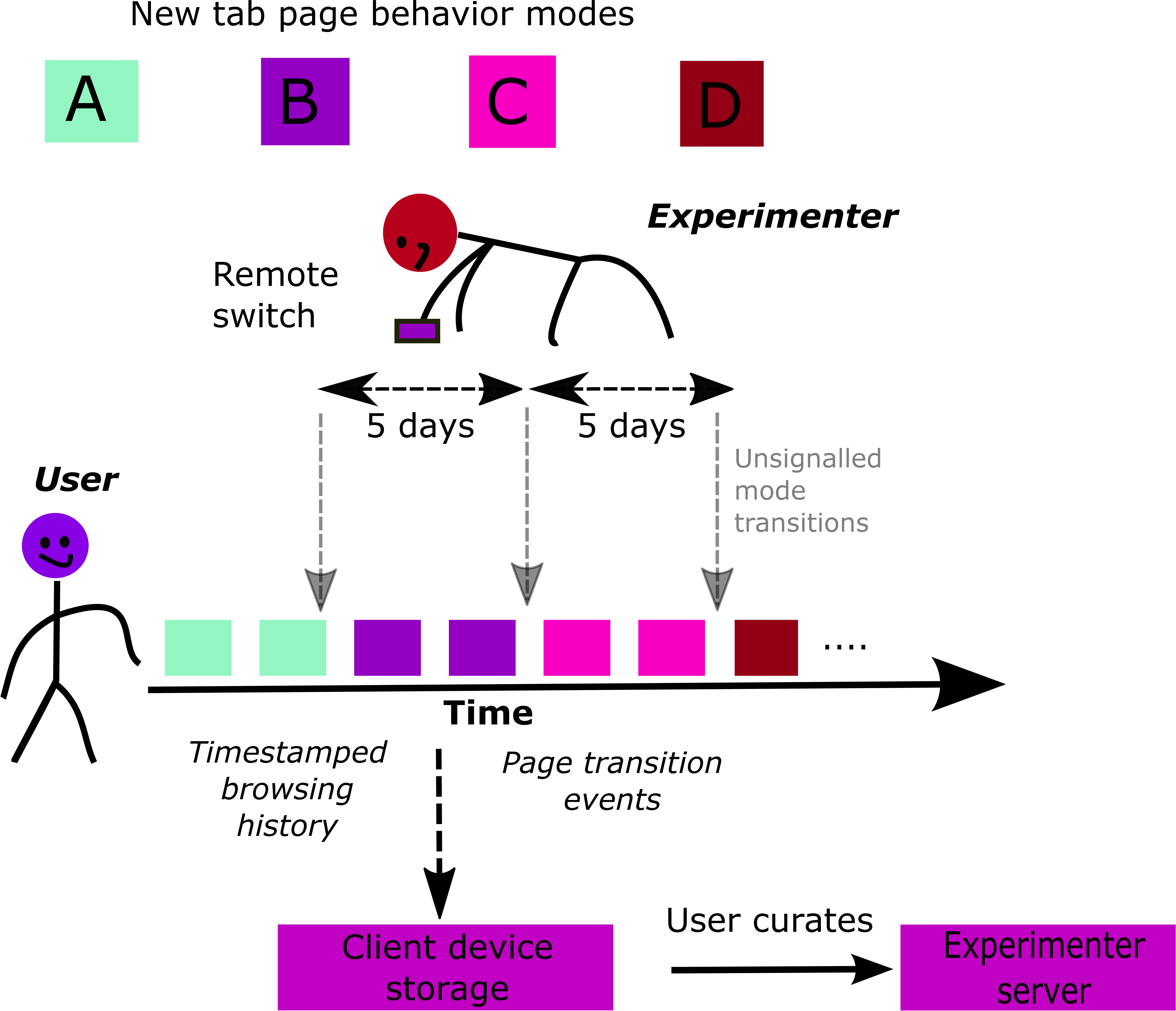}
\caption{The experiment design}
\label{fig:design}
\end{figure}
\end{center}

We designed a Firefox web extension using JavaScript, HTML and CSS.  It broadly consists of three parts, a webpage to replace the default new tab display, a browser action page to log the browsing data into a local file and a remote script located in the experimenter's server that is used to tweak the type of websites displayed in the new tab page, as shown in Figure \ref{fig:design}. In this design, the experimenter does not have access to the participant's data during the course of the experiment. At the end of the experiment, the experimenter explicitly asks for the logs, and participants are advised to curate their history to remove personal and private content before handing them over, if they choose to do so. 

We programmed four different behavior modes in the replacement new tab page generated when people install our extension, controlled by a parameter in the extension script that we could change remotely without the user's knowledge. Each participant's replacement new tab page shows the web page icons as per exactly one of the following four modes at a given time. 
\begin{enumerate}
\item Display the \textbf{most} visited sites in browsing history
\item Display the \textbf{least} visited sites in browsing history (at least one visit, naturally)
\item \textbf{Default} behavior, following Mozilla's frecency algorithm - a combination of frequency and recency~\cite{frecency}
\item Display a \textbf{blank} page
\end{enumerate}

Our web-extension logs the webpages the user visits, the time of visits and the transition type from one webpage to the next - {\bf typed} in the address or search bar, {\bf linked} to from the content of a different site, or {\bf clicked} through the browser UI. It also automatically logs all the websites currently displayed in the new tab page whenever it is opened. The replacement new tab page is designed to visually resemble the default Firefox new tab page so that after a certain period of time, the user is not very conscious of the extension running in background. Once the extension is installed on a participant's computer, the experimenter can change the behavior of the replacement new tab page by simply changing a parameter in a participant-specific file hosted on the experimenter's server. We used a fixed switching interval of 5 days, with all participants beginning with the default mode of new tab behavior, and with further switches assigned pseudorandomly, ensuring that all participants received roughly equal duration of exposure to each of the three new behavior modes.  Since we used a rolling recruitment process the 5-day blocks don't systematically interact with weekends for all participants, an important consideration in a behavior measurement experiment. This interaction is approximately random across subjects in our experiment. We used a pseudo-random allocation of conditions instead of using a fixed schedule counterbalanced across participants because in our in-the-wild study, we could not control when participants would stop participating in the experiment, and expected large attrition rates. If all participants completed the full term of the experiment, we would have approximately the same block-wise assignment as if we had assigned fixed blocks to begin with. By good fortune, we had a low attrition rate (~20\%), and all eighty participants who lasted the full two month term complied with our reminders for sending in browsing logs immediately. As a result, condition viewing time was approximately balanced both across and within our sample, averaging {24.46, 25.18, 24.56, 25.78} by percent across our four conditions (Most visited, Least visited, Default, Blank UI) with standard deviations {1.41, 1.81, 1.49, 2.41}  across all 80 participants. Further, rolling recruitment randomizes weekday-weekend interactions among participants.

\subsection{Sample}
Eighty volunteers from the general population (32 female, age $28.2 \pm 6.1$ years), recruited on a rolling basis, consented to participate in our experiment. 30\% of them were university students, 40\% were employed in the industry (excluding software industry), 20\% were employed in the software industry and 10\% were employed in the academia (lecturer/post-doc/professor). There were no web-developers in the sample, who used the browser for development. Participants were recruited through advertisements in newspapers, on social media websites and emails. Participants were not incentivized and were naive to the purpose of the study. We explained to them that we were interested in analyzing their web browsing history for a research experiment using a software extension they would install into their browser, that our extension would only store their data locally, and that they would have complete control over the curation process that would determine what part of their history they would share with the experimenters. The participants used various devices including laptops, desktops, tablet computers and mobile phones and no restriction on the type of device for a participant was imposed. A total of 10 participants reported to have used mobile phones / tablet computers while the remaining participants used either a laptop / desktop computer for installing the extension. At the end of two months of browsing for each participant, we asked them to curate the log files and share them with us, retaining the option to refuse. The two-month long browsing logs volunteered by these eighty participants formed our primary dataset. 42 of the 80 participants continued using the web extension for a longer duration (another two months). The results from this additional data are presented in Section 3.1. 

At the time of submission of the logs, we asked the participants what fraction of the website logs had been curated by them. None of the eighty participants had curated more than 5\% of the total content, with most participants not curating at all, which is reasonable and does not negatively affect our study. We had not disclosed to these participants that we intended to manipulate their new tab page's behavior remotely. To assess the extent to which they may have become aware of this possibility,  at the time of collecting the logs we also asked them whether anything about the websites displayed in the new tab page `bothered' them over the past two months.  They were asked to express their opinion on a $0$-$10$ scale with $0$ being the `not aware' state and $10$ the `very aware' state. They were not specifically asked if they thought the type of websites shown were being altered remotely to avoid experimenter demand effects~\cite{charness2012experimental}. Subjects' responses are shown in Table~\ref{tb:aware}.

\begin{table}[tbhp]
\caption{Participants' level of awareness of our experimental manipulation.}
\begin{center}

\begin{tabular}{ |c|c|c|c| }
 \hline
 Opinion & 0-4 & 5-7 & 8-10 \\
 \hline
 Number of subjects & 70 & 7 & 3 \\
 \hline
\end{tabular}

\label{tb:aware}
\end{center}
\end{table}

These survey results suggest that the majority of our participants remained naive to the purpose of the experiment, and continued to behave naturally in the face of our remote manipulations. 

\subsection{Data processing}
The self-curated browsing logs obtained from all 80 participants in the experiment constituted our primary analysis material. These logs contained time-stamped instances of different website URLs visited by the participants. From these logs, we filtered out pop-ups and ads using a white-list, and used a 30-minute gap between page views to delimit browsing sessions. We collapsed web-pages visited on the same primary web domain into multiple visits to the same website, and likewise collapsed page refreshes (immediate repetitions of the same page in the log) as a single visit to the page. The few occurrences of dynamic web-pages in the browsing log were collapsed to that of the parent domain. We explicitly handled pages visited from search engines in the browser. For all occurrences of Google, Yahoo, Bing, Ask.com, Baidu, Wolframalpha and DuckDuckGo in the browsing log, we considered the page clicked immediately afterwards in the search results. If the search engine address was typed out in the new tab page, we treated the page clicked from the search results as if it was typed out in the address bar.

\section{Results}

Since our hypothesis generally concerns the potential reduction of the diversity of websites that users visit as a consequence of the format of the new tab page display, our first coarse measurement of this property was the number of unique websites (defined via the primary web-domain in each URL) visited by our experiment participants during each mode of the new tab page's display properties. 
\begin{center}
\begin{figure}[htbp]
\includegraphics[width=0.95\textwidth]{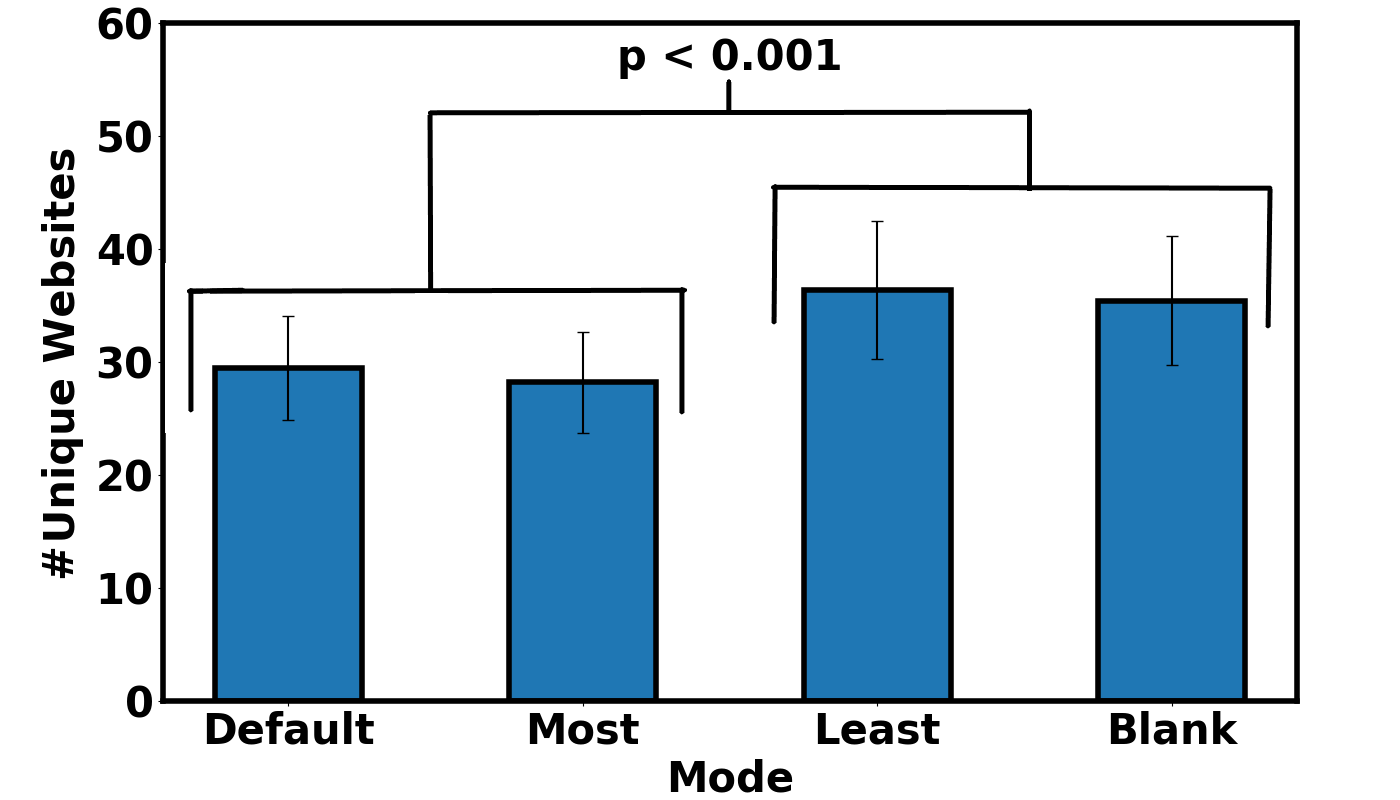}
\caption{ Number of unique website count for each of the four modes averaged over all 80 individuals. Error bars represent $\pm$ 1 Standard Deviation (S.D.).}
\label{fig:pairwise}
\end{figure}
\end{center}
Combining the default and most visited modes' data into one sample and for the blank and least visited modes' data into another, a two-sample T-test yields strong statistical significance  $p < 10^{-3}$ and a very large effect size (Cohen's d = 1.81) as shown in Figure~\ref{fig:pairwise}. Thus, there is clearly a large difference in user behavior across these experimental cohorts, in the predicted direction. Users visit a lot more unique websites on average when the new tab page is blank or shows infrequently visited websites, suggesting that the presence of recommendations by default is suppressing the repertoire of webpages they might naturally visit. For completeness, we show the results of bonferroni corrected pair-wise  T-test among all the pairs of conditions in our study (Table~\ref{tb:pairwise}).
\begin{table}[tbhp]
\caption{Results of Pairwise T-test for all the pairs of conditions (as shown in Fig. 3). Number of unique website counts for each of the four modes are averaged over 80 individuals. }
\begin{tabular}{@{}llll@{}}
\toprule
\textbf{Condition 1} & \textbf{Condition 2} & \textbf{p-value} & \textbf{Inference}         \\ \midrule
Most  visited        & Least visited        & 0.000011         & \textbf{p \textless 0.001} \\
Most visited         & Default UI           & 0.620102         & p\textgreater{}0.05        \\
Most visited         & Blank UI             & 0.000008         & \textbf{p\textless{}0.001} \\
Least visited        & Default UI           & 0.000024         & \textbf{p\textless{}0.001} \\
Least visited        & Blank UI             & 0.522001         & p\textgreater{}0.05        \\
Default UI           & Blank UI             & 0.000018         & \textbf{p\textless{}0.001} \\ \bottomrule
\end{tabular}
\label{tb:pairwise}
\end{table}
However, the heterogeneity of web browsing behavior across individuals could conceivably inflate these statistics, in the sense that a few highly prolific web users in our participant pool could skew the cohort-level differences should they happen to conform to our hypothesis. Thus, it is important to also analyze the data at the individual-level. We plot the difference between behavior during blank UI mode and default mode for all participants in Figure \ref{fig:inds}. Note that not a single of the eighty participants visited more unique websites while the new tab page behaved under its default settings, vis-a-vis when the default recommendation display was switched off (Blank UI mode). A one-sample T test of these absolute differences with respect to zero sample mean was statistically significant at $p < 10^{-6}$. The average participant's unique site visit count increased by 15\% (median improvement 12\%) when the new tab page was set to display a blank page over their corresponding counts when the new tab page worked by default.
\begin{center}
\begin{figure}[htbp]
\includegraphics[width=0.95\textwidth]{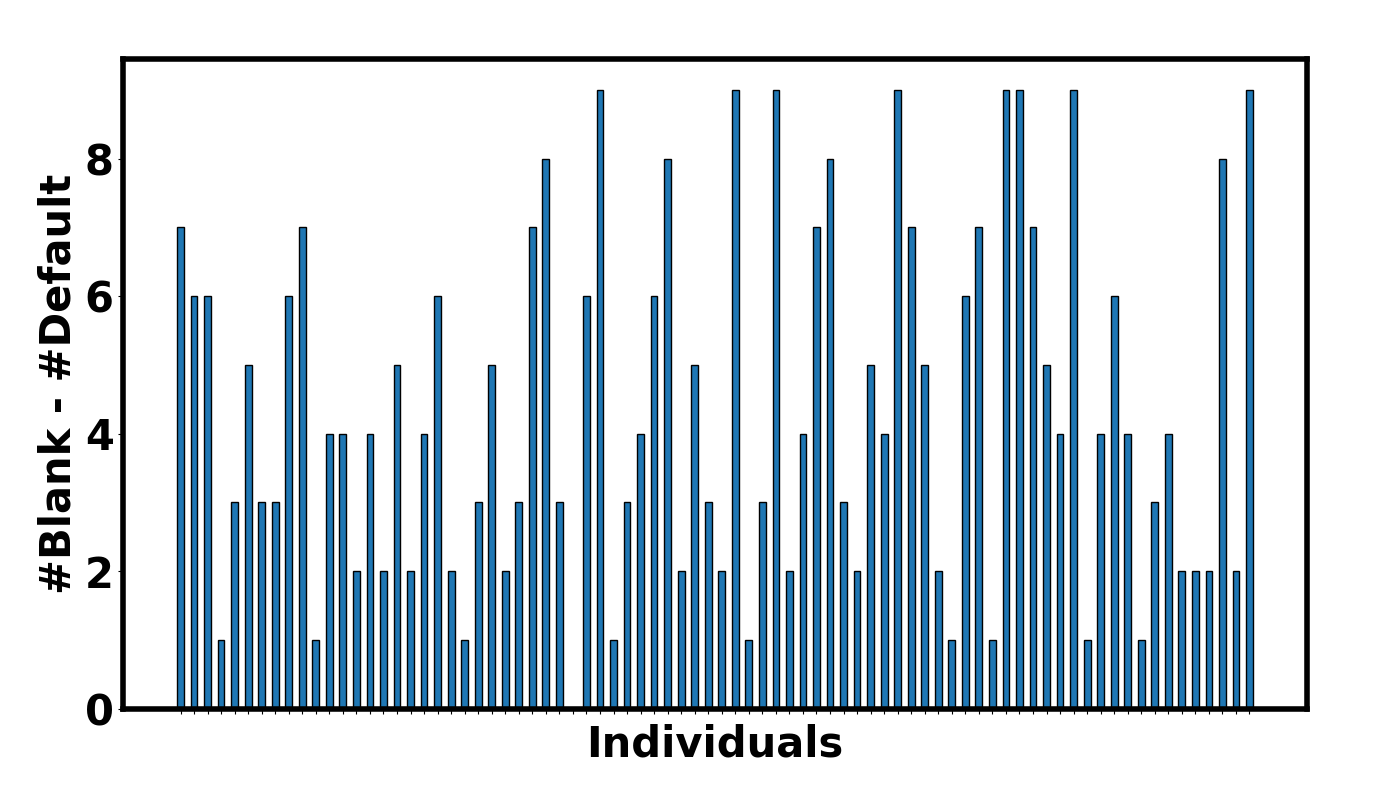}
\caption{Absolute difference between the number of unique websites visited under the blank new tab page mode and under the default mode for all experiment participants.}
\label{fig:inds}
\end{figure}
\end{center}
It is reasonable to conclude from this observation that changing the behavior of the new tab page does in fact reduce the diversity of websites browsed. But this analysis does not substantiate our hypothesis about the mechanism responsible for this reduction in diversity - a reduction in users' propensity to explore content, caused by interaction with the new tab page interface.  

\begin{center}
\begin{figure}[htbp]
\includegraphics[width=0.95\textwidth]{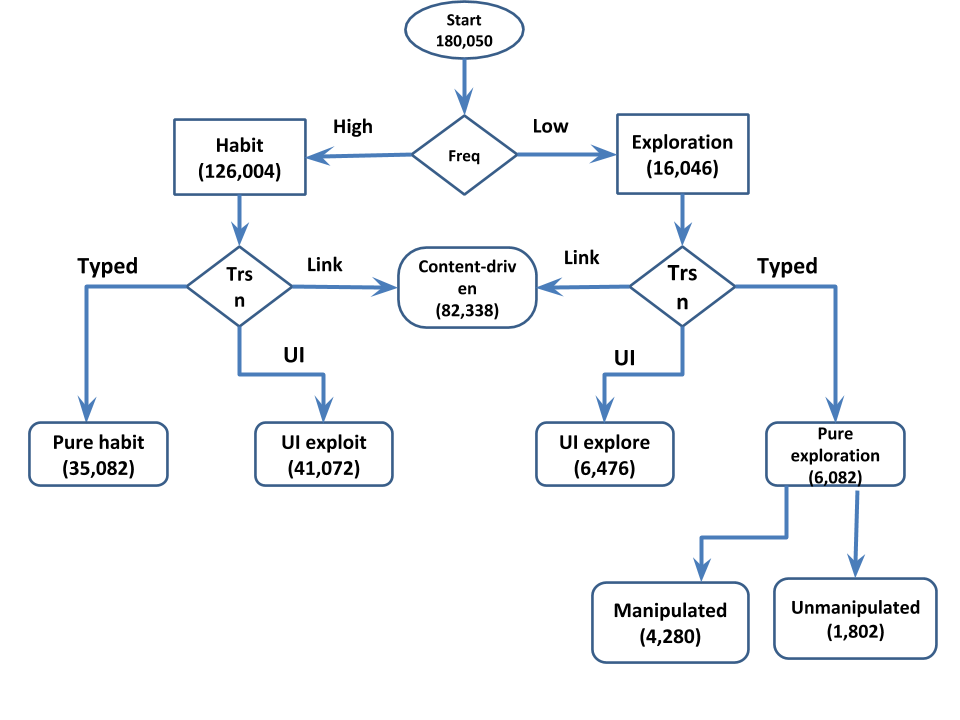}
\caption{The analysis flowchart. The two decision points (diamond boxes) represent decisions by frequency (high = 4th quartile, low = 1st quartile) and transition type (link, typed, UI). Numbers inside boxes represent the number of webpage visits categorized under each category.}
\label{fig:venn}
\end{figure}
\end{center}

We needed a more granular view of the data to see if this might actually be true. Recall that we store both the timestamped page visit, and the web event that brings the user to that page. We categorized the pages according to the visit frequency and transition type to focus especially on page visits wherewith the user's {\em exploratory} intent to retrieve information from a new information source can be discerned. This categorization is illustrated with a pseudo-flowchart in Figure \ref{fig:venn}, showing visually how we use the two decision points (frequency and transition type) to categorize web page visits into non-exclusive categories. To interpret the set-theoretic implications of the flowchart correctly, both {\em UI} and {\em pure} exploration are subsets of exploration, but {\em exploration} and {\em habit} are mutually exclusive. A brief description of these terms and their rationale follows.

\vspace{-0.4cm}
\hspace{0.3cm}

\begin{itemize}
\item {\bf Habit.} Websites with high visit counts (removing the influence of interstitial URLs) are expected to correspond to content that users are in the habit of accessing. Since the number of webpages an individual visits will inform what constitutes high for them, we categorize as {\em habit}-based visits all page visits to webpages whose visit frequency occurs in the fourth  quartile ($Q_4$) of visit frequencies sorted in ascending order per browsing history log. In a measure of the heavy skew in the distribution of webpage counts, this quartile accounts for about 70\% of all site visits across all participants' data. The absolute value of average visit count of websites in the fourth quartile ($Q_4$) across all 80 participants is $42\pm 8$. That is, on average, participants visited websites in this category thrice in every four days during our 60 day observation period. A distinct sub-category ``pure habit" takes into account frequently visited webpages visited by typed transitions from the address bar (or search the web box) of an existing tab.

\item {\bf Exploration.} As above, it is intuitive that when people are browsing in an exploratory manner, such browsing will be marked by visiting new sources of content, hence new webpages. We mark as {\em exploration}-based visits all visits to pages that fall in the first quartile ($Q_1$) of visit frequencies sorted in ascending order per browsing history log\footnote{Not all visits to infrequently visited webpages constitute exploration. Sometimes users may have only need to go to a website, like a bank website very infrequently, on purpose. Thus, our definition overestimates the amount of exploration. Since we are ultimately interested, as we will see below, in the change in the amount of exploration under controlled manipulation, this overestimation of the absolute quantity ends up not being crucial.}. This quartile accounts for about 9\% of all site visits across all participants' data. The absolute value of average visit count of websites in the first quartile ($Q_1$) across all 80 participants is $8\pm3$. That is, on average, participants visited websites in this category about once a week during our 60 day observation period. 

\item {\bf Content / Event -driven.} Since we are interested in the interaction between the user's mind and the browser interface, we want to remove from our analysis all webpage visits that originate from an information source outside of these two. To this end, we mark all hyperlink based visits (marked in our logs as having the transition type `link') irrespective of their frequency rank as content-driven, and exclude them from our analysis.  

\item {\bf UI exploitation.} The presence of clickable page icons obviously makes those webpages more accessible to users, over and above their mental propensity to visit these sites. Websites that users visit frequently, but use the UI to access, constitute an interesting sub-category of the {\em habit} category. For these webpages, it is not as clear as for others that repeated visitation is purely a function of a hedonic preference to visit them. The increased accessibility of the pages is clearly, but unquantifiably, also a factor. We refer to this category of pages, selecting all habit pages arrived at either via UI clicks or by typing in the URL bar while the specific page icon was visible in the new tab page, as {\em UI exploit} pages.

\item {\bf UI exploration.} When users visit rarely visited webpages via interaction with UI elements, we can infer that the UI interaction is driving them away from their typical behavior.  This makes the subset of {\em exploration} pages that users arrive at from the new tab page (whether with transition \lq link \rq or \lq typed \rq), especially interesting. We call this sub-category {\em UI explore}.

\item {\bf Pure exploration.} The final category we define is, expectedly, sparsely populated, representing about 3\% of all web page visits in our dataset. We define this as the subset of {\em exploration} pages whose icons are not displayed on the new tab page, have a transition type of `typed' and have been transitioned to directly from the new tab page. To be a member of this set, the user has to have entered the webpage URL (at least its first few letters, keeping in mind auto-complete capabilities of browsers) with no input from the browser UI. Visiting rarely visited sites without the (measured) influence of either other websites' content or the content of UI elements means that this subset of webpage visits reflect exploration driven purely by memory considerations. Privileging the user's mental content in a semantic sense, we label this category {\em pure exploration}. Within this category, we make two further distinctions: {\em manipulated} pure exploration takes into account only typed transitions to rarely visited pages from the new tab page. {\em Un-manipulated} pure exploration takes into account typed transitions to rarely visited pages from the address bar (or web search box) from an existing tab.

\end{itemize}

\begin{figure*}[htbp]

\begin{center}
\includegraphics[width=1\textwidth]{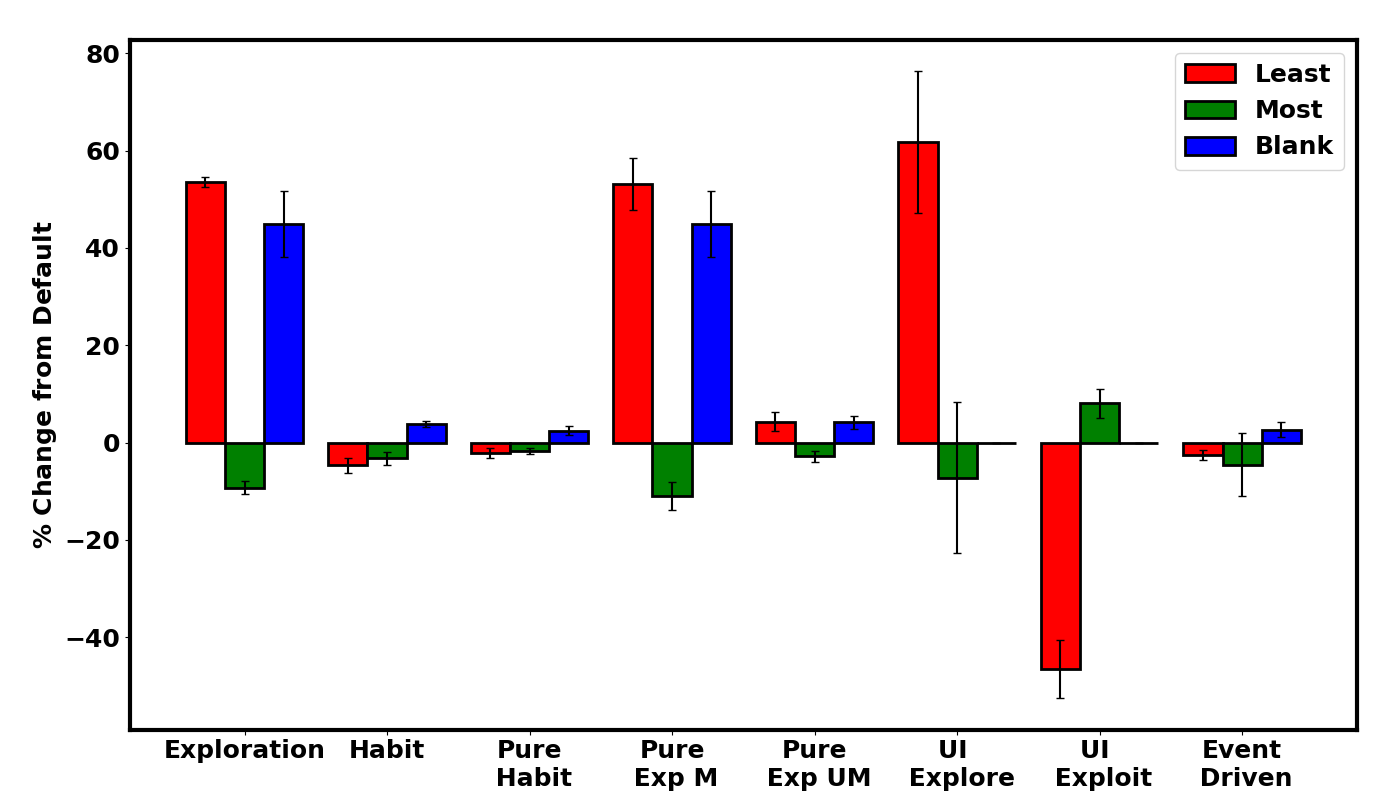}
\end{center}
\caption{The change in each category's occurrence fraction with respect to the same fraction seen in the default mode per subject, averaged across all subjects. Error bars represent $\pm 1$ S.D. Pure Exp UM and Pure Exp M respectively denote Pure Exploration Un-manipulated and Manipulated.}
\label{fig:cohort}
\end{figure*}

Using this categorization of browsing events, we sought to quantify the category-wise change in users' web browsing behavior across the three different UI manipulations. Because the different categories defined above have vastly different base rates, it is most sensible to measure the change across UI modes in terms of the category-wise percentage change, 
\[100\times\left(\frac{\text{New- category- occurrence- frequency}}{\text{Default- category- occurrence -frequency}} - 1\right),\] where the category occurrence frequency is mode-sensitive and is simply the number of page visits marked in a particular category for a user divided by their total page visits while their browser's new tab page's behavior was {\em in a particular mode}. 

Figure~\ref{fig:cohort} summarizes the variation in this quantity as a function of mode manipulations, averaged across all subjects for all categories defined above. It is evident that behavior is not changed, in aggregate, as a consequence of these manipulations. The bulk of page visits in our dataset are tagged as {\em habitual} visits, and the ratio of habitual site visits changes insubstantially if at all across our users. Likewise, for content-driven behavior, which shows no statistically meaningful change. These null findings are reassuring. They suggest that our manipulations did not overtly change subjects' browsing behavior by introducing experimenter demand effects~\cite{charness2012experimental}. This is also congruent with the survey results of low awareness of our manipulations shown in Table~\ref{tb:aware}. 

If the presence of web page recommendations on the new tab page affects behavior by reinforcing habitual browsing patterns, we expected to see that our exploration-linked categories would show positive percentage changes with respect to the default condition when we remove all page icons from the new tab display (the `empty UI' condition) and when we present users with their least visited webpages as icons in the new tab display (the `least visited' condition). As Figure \ref{fig:cohort} shows, this prediction is borne out clearly by our data. For both these conditions (least visited and empty UI),  exploration and pure exploration categories showed large positive changes ($\sim50\%$) with respect to the baseline condition (the default frecency-based UI web page icons display). These conclusions are statistically significant in Bonferroni-corrected pairwise T-tests ($p<0.001$), with effect sizes \{0.83,0.89\} for exploration and \{0.87, 0.85\} for pure exploration, for the least visited and blank UI display with respect to the default condition respectively. 

Further, note also in Figure \ref{fig:cohort} that pages in the {\em manipulated} pure exploration category accounted for almost all the change in the pure exploration category as a whole; there are only marginal increases in the {\em non-manipulated} pure exploration category. The difference between these two categories was only that the manipulated category page visits started with typing in URLs while the new tab page recommendations were visible, whereas the non-manipulated category page visits started from other pages, with new tab recommendations not visible. Since all other aspects of the users' experience were constant across these two sub-categories, the stark difference in browsing behavior emerged only by virtue of the new tab page recommendations being visible or invisible. This performance dissociation allows us to confidently pinpoint the source of the difference in browsing behavior - the recommendations on the new tab page - with precision.



Secondary observations include an interesting pattern seen in the difference between the default condition, which uses Mozilla's frecency algorithm~\cite{frecency}, and a simple approximation of it that uses only frequency information, ignoring recency - our `most visited' condition. We find a substantial negative impact on exploration for this condition with respect to the frecency condition, statistically significant ($p<0.001$). This suggests that, by accommodating the role of temporal context, frecency supports users' expression of their naturally diverse preferences more than context-insensitive pure frequency-based displays. Since we did not seek to differentiate these two conditions in our experimental design, this conclusion is necessarily speculative. Further research is needed to reproduce and substantiate this finding.

Secondary observations also include a pair of common-sense findings that UI explore increases and UI exploit decreases in the `least visited' condition. This is mostly because the `least visited' manipulation makes sites tagged `explore' accessible and sites tagged `exploit' inaccessible to UI interaction. This supports the view that accessibility heavily influences users' browsing choices, in line with similar effects in search and social media selections, but our present experimental design is unable to quantify the size of this effect with any greater precision.

\begin{figure}[htbp]
\centering

\begin{tabular}{cc}

& \textbf{Exploration} \\ 
 
&      \includegraphics[width=.80\textwidth]{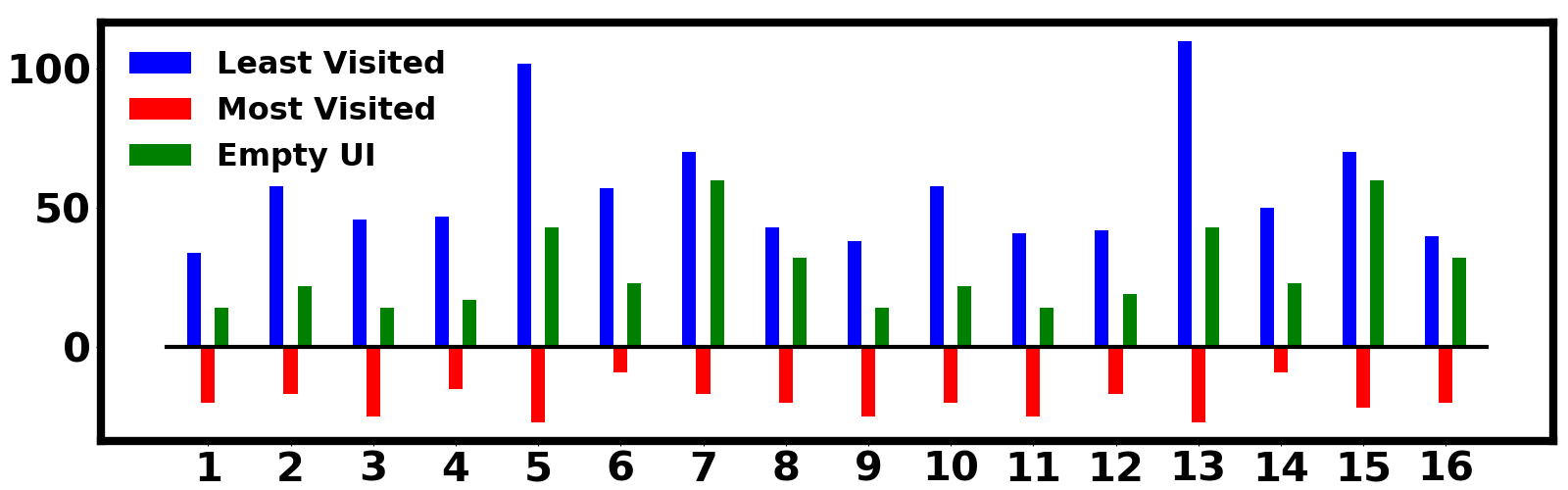}\\ 
    \multirow{4}{*}{\begin{turn}{90} $\%$ \textbf{\Large Change from default behavior} \end{turn}}
                  &   \includegraphics[width=.80\textwidth]{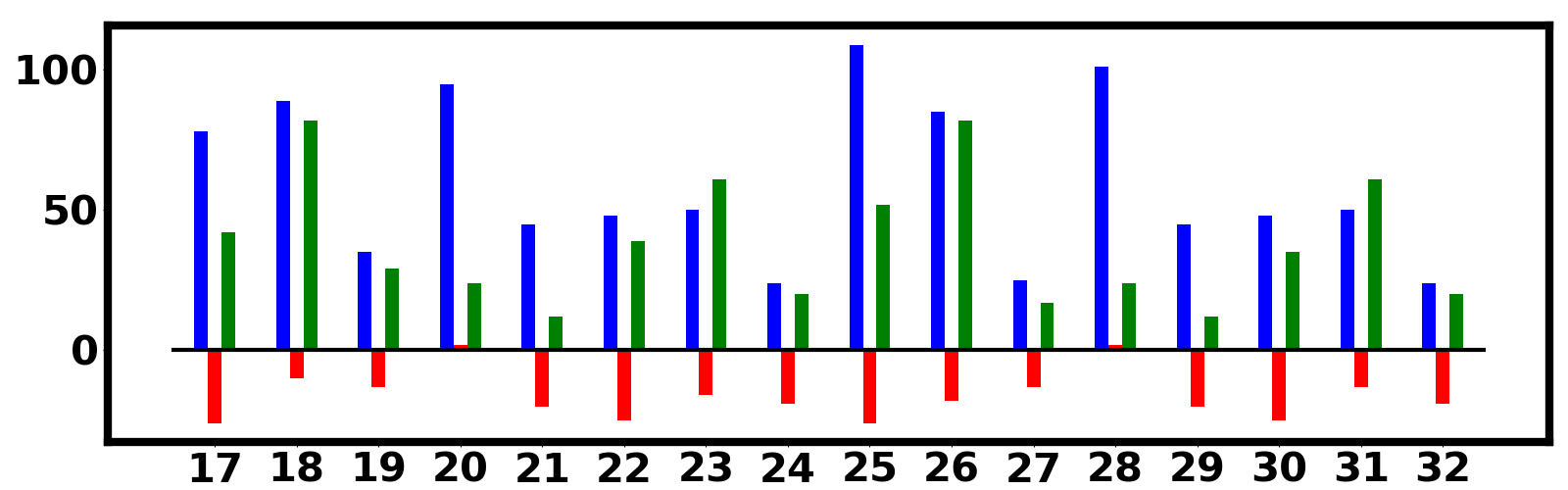}\\                   
                  &         \includegraphics[width=.80\textwidth]{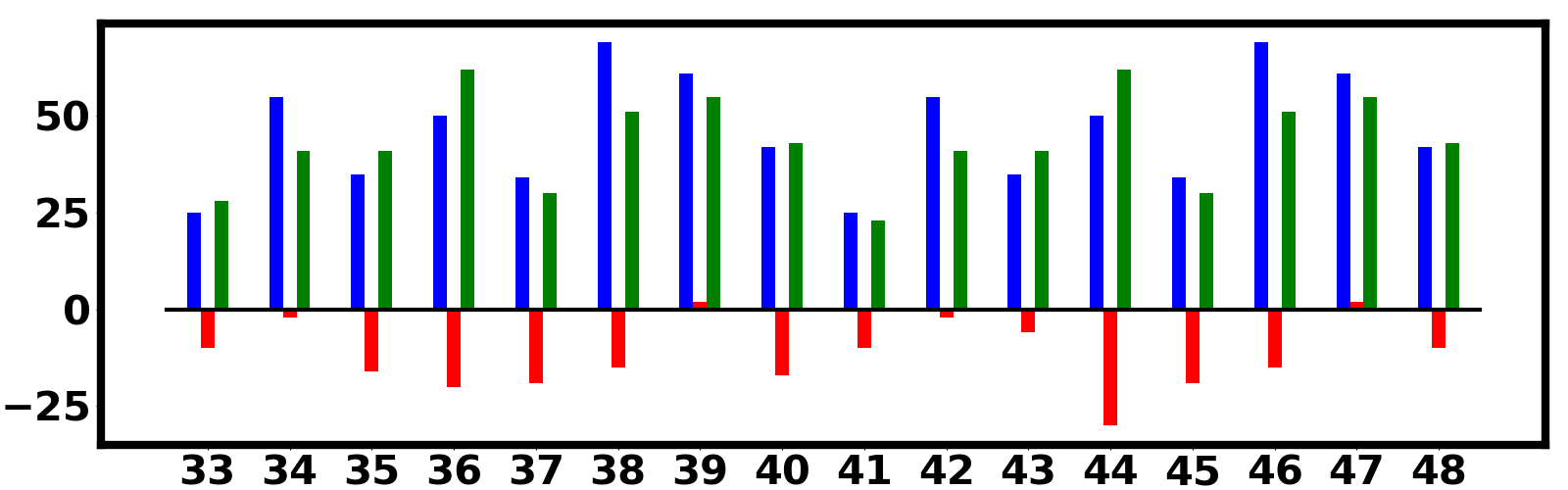} \\               
                  &         \includegraphics[width=.80\textwidth]{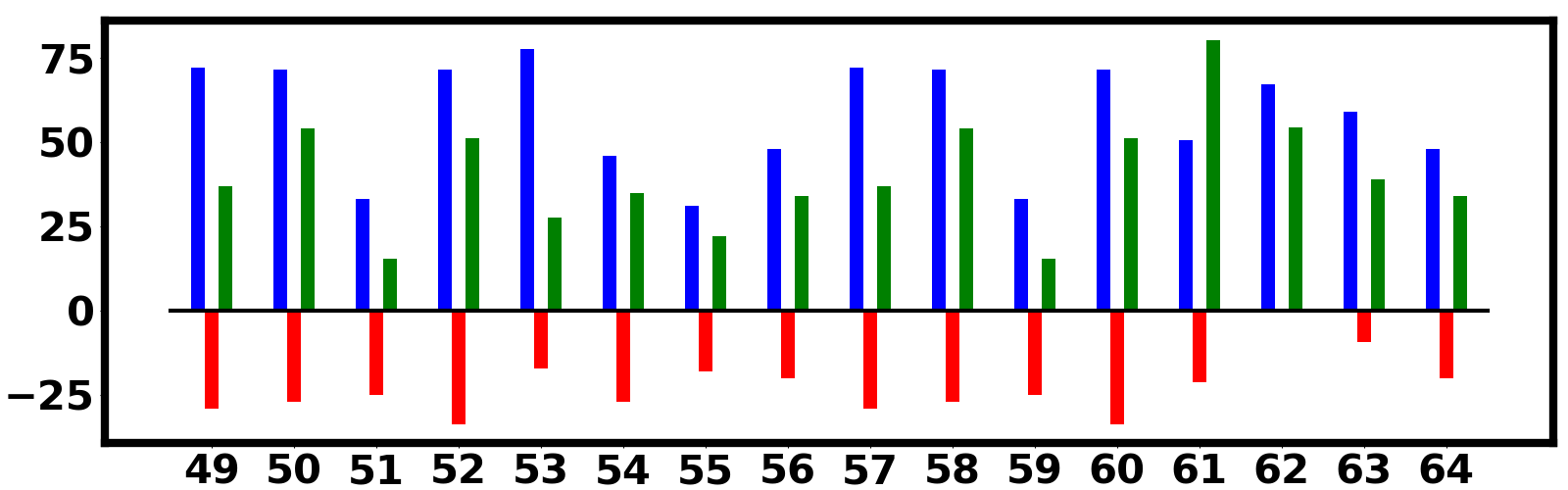} \\
     &\includegraphics[width=.80\textwidth]{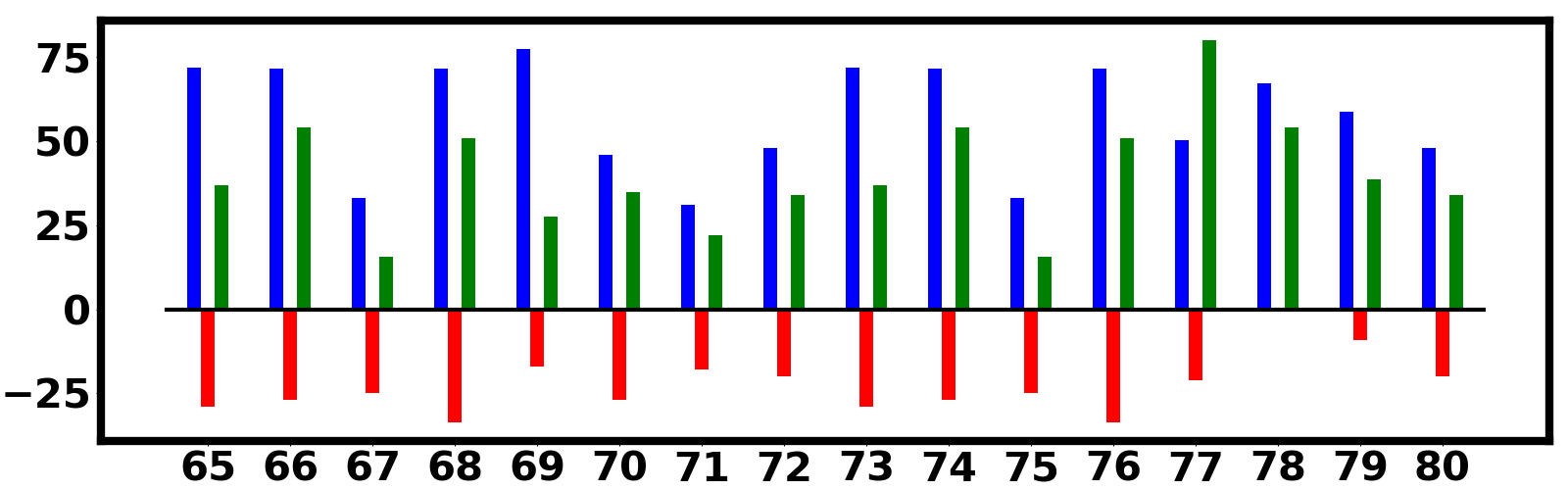}\\
                  & \multicolumn{1}{l}{} 
\end{tabular}
\caption{Pattern of percentage change for overall exploration behavior for all the eighty individual subjects. Legends in top panel propagate through to all panels.}
\label{fig:individuals}
\end{figure}

\begin{figure}[htbp]
\centering

\begin{tabular}{cc}

& \textbf{Pure Exploration} \\ 
 
&      \includegraphics[width=.80\textwidth]{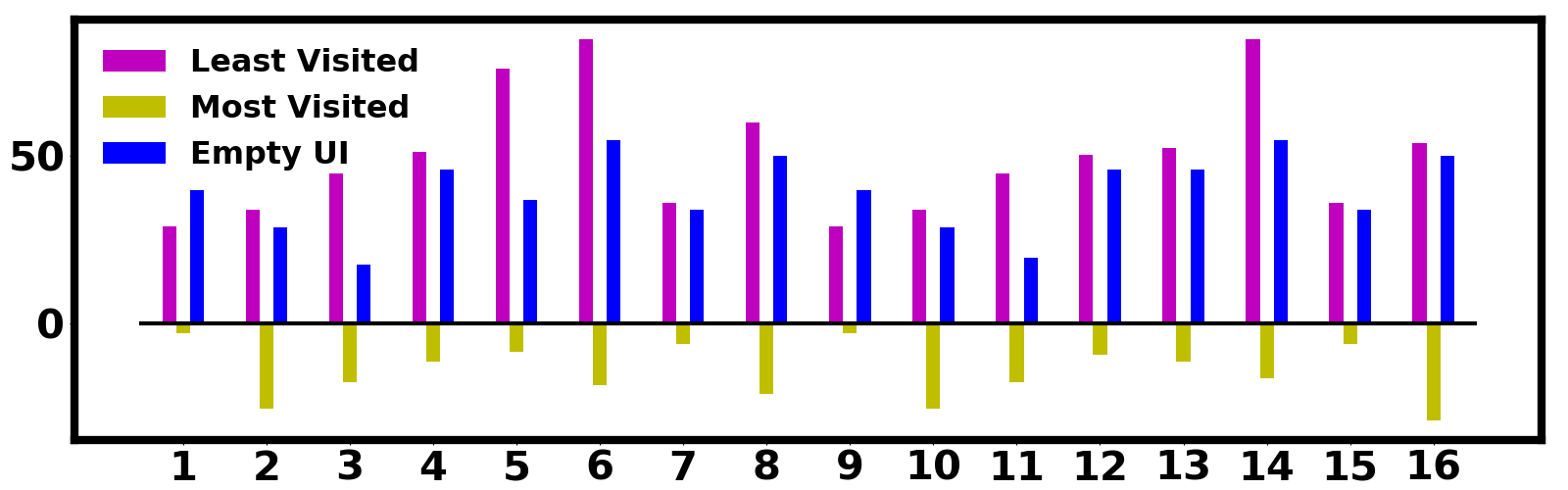}\\ 
    \multirow{4}{*}{\begin{turn}{90} $\%$ \textbf{\Large Change from default behavior} \end{turn}}
                  &   \includegraphics[width=.80\textwidth]{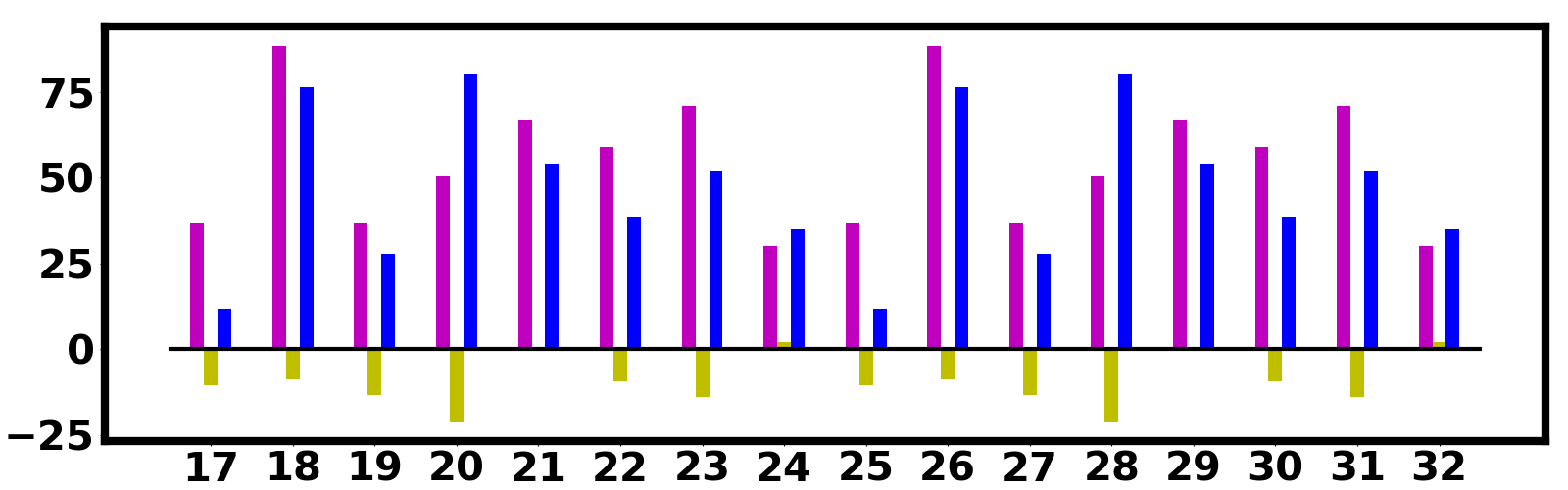}\\                   
                  &         \includegraphics[width=.80\textwidth]{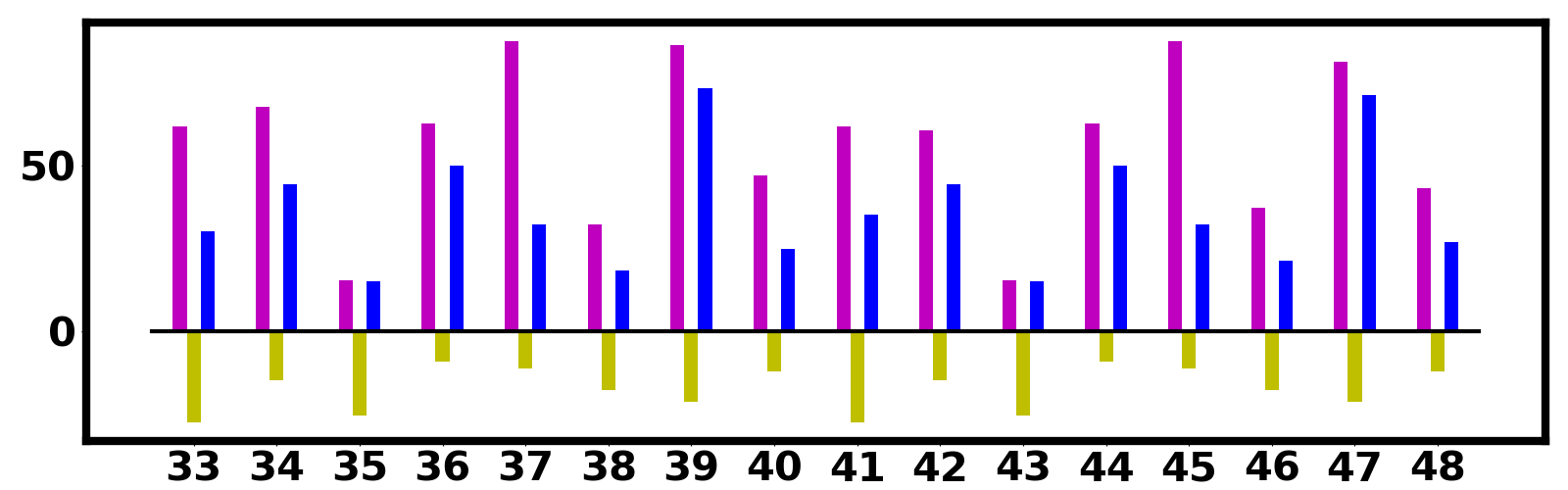} \\               
                  &         \includegraphics[width=.80\textwidth]{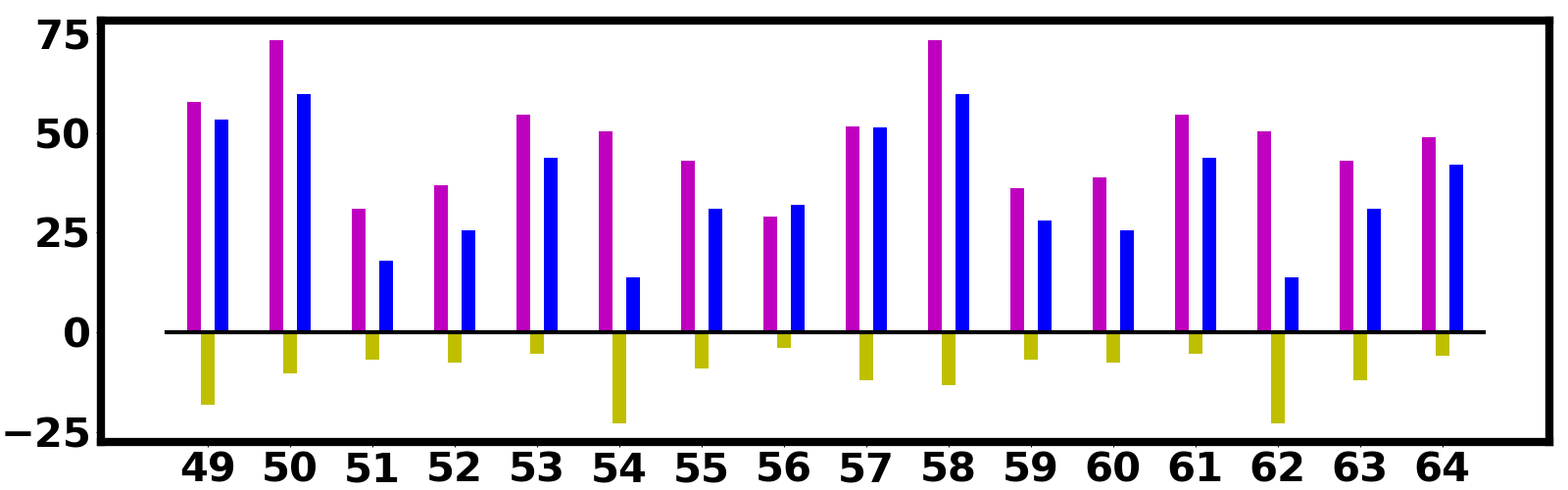} \\
     &\includegraphics[width=.80\textwidth]{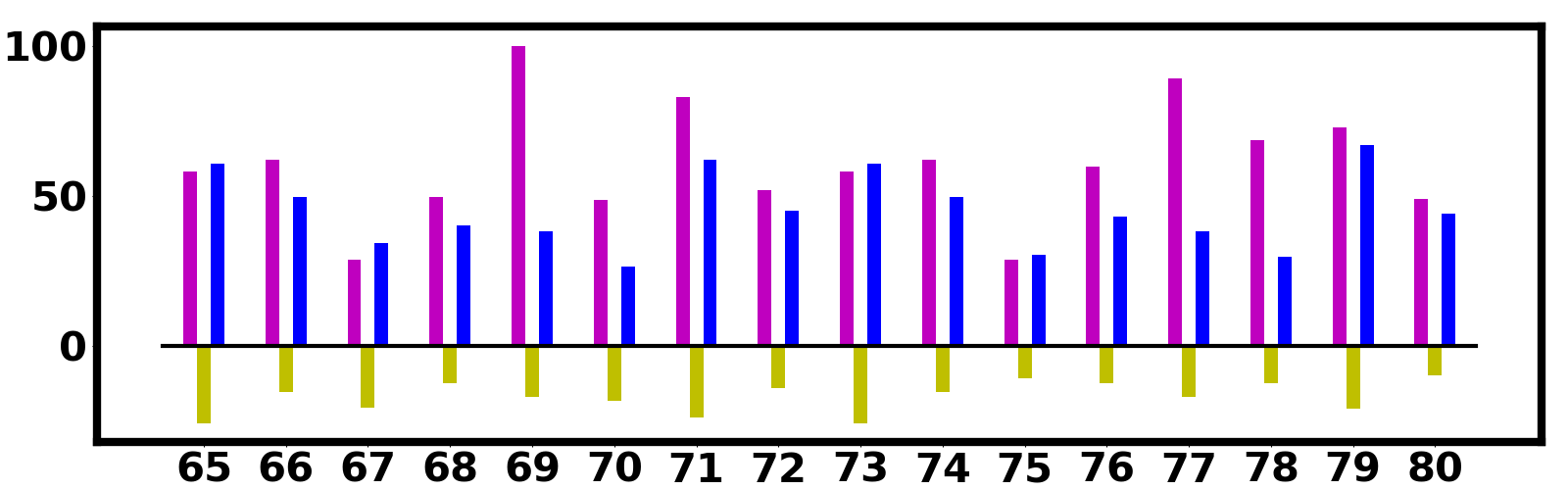}\\
                  & \multicolumn{1}{l}{} 
\end{tabular}
\caption{Pattern of percentage change for overall pure exploration manipulated (Pure Exp M) behavior for all the eighty individual subjects. Legends in top panel propagate through to all panels.}
\label{fig:individuals2}
\end{figure}

Our experimental results justify the primary hypothesis we wanted to test - that manipulations of the UI in the direction of presenting no or infrequently visited page icons would increase users' propensity to visit webpages they are not habituated to visiting {\em of their own volition}. However, this is a statistically marginal effect when observed across the entire span of behavior captured in our browsing logs. No more than 1\% of all page visits are categorized as pure exploration, so even large relative changes in this category's representation are small shifts in user behavior in the aggregate. Thus, a more convincing analysis of the data would try to look for evidence beyond the purely statistical to show that the effect we are postulating is real.

One way to do this is to see if the pattern of behavior we see at the cohort level also shows up in individual subjects. Our hypothesis is a hypothesis about individuals, not populations. So if the results we are seeing are genuine and not the result of statistical fluctuations in the aggregate database, the same pattern of increased exploration for the least visited and empty UI conditions and reduced exploration for the most visited UI condition should hold across most of our subjects' individual data. Figures~\ref{fig:individuals} and~\ref{fig:individuals2} show the percentage change in exploration and pure exploration for each of our 80 individual subjects. It is immediately evidence that the within-individual changes across the three conditions closely match the averaged changes observed across the sample. This lends further credence to the view that, while subtle in aggregate, the influence of the UI manipulations affects user psychology at a personal level precisely in the directions we have predicted on theoretical grounds.

An interesting observation arises from the participants who reported to using their mobile phones (smartphones) / tablet computers for the experiment. These were the participants numbered 10, 20, 26, 33, 34, 44, 68, 77, 78 and 79 . Rest of the participants used either a laptop / desktop computer for installing the extension we provided. As evidenced by Figures~\ref{fig:individuals} and~\ref{fig:individuals2}, their $\%$ change in exploration and pure exploration from the default mode to the `least visited' and `blank UI' modes is quite high. In particular, the average change over these six individuals is around 10$\%$ higher for the `least visited' mode and around 14$\%$ higher for the `blank UI' mode for exploration over the aggregate for the entire sample shown in Figure~\ref{fig:cohort}. One of the reasons for a more acute suppression of exploration in the default mode for mobile device users could be due to the smaller screen sizes that incentive clicking on the web-page icons of the new tab page over typing in the address bar. However, since the number of individuals in our sample who used a mobile device is very low, additional experimentation is needed to accurately quantify this distinction. 

\begin{figure}[htbp]

\begin{center}
\includegraphics[width=0.95\textwidth]{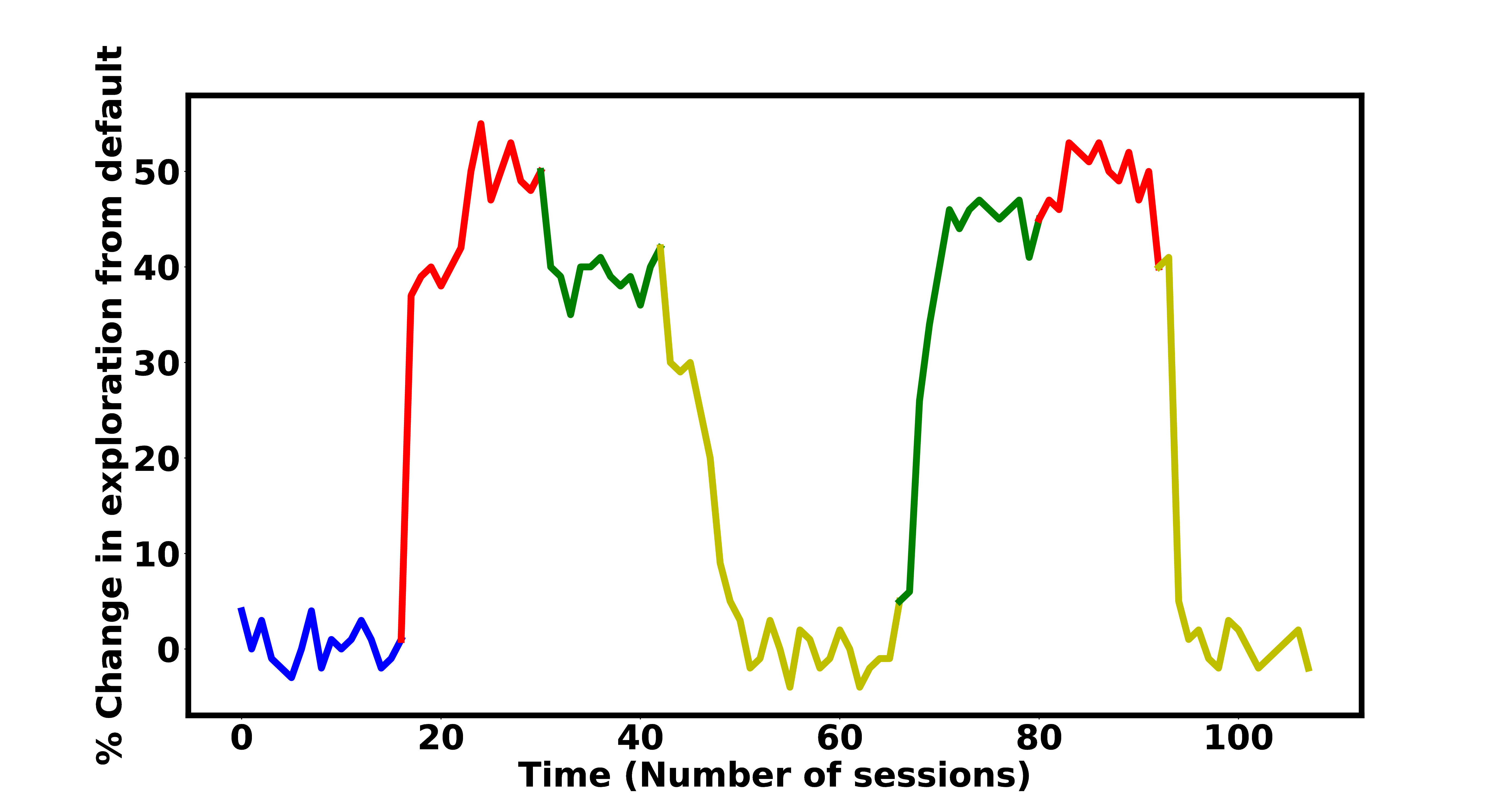}
\end{center}
\caption{Temporal variation of {\em exploration} for one participant. The y-axis shows the $\%$ change in exploration from the averaged (across sessions) value in Default UI mode. The x-axis shows the flow of time in number of sessions since the beginning of experiment. ({\em Blue - Default, Red - Least Visited, Green - Blank UI, Yellow - Most Visited})}
\label{fig:temporal}
\end{figure}

\begin{table}[tbhp]
\centering
\begin{tabular}{@{}ccccc@{}}
\toprule
                       & \textbf{\begin{tabular}[c]{@{}c@{}}Default\\ Sites\end{tabular}} & \textbf{\begin{tabular}[c]{@{}c@{}}Least \\ Visited\end{tabular}} & \textbf{\begin{tabular}[c]{@{}c@{}}Most \\ Visited\end{tabular}} & \textbf{Blank UI} \\ \midrule
\textbf{Default sites} &X & 41.08 $\pm$ 4.4 & -0.88 $\pm$ 0.1& 33.15 $\pm$ 3.2\\ 
\textbf{Least Visited} &-36.68 $\pm$ 3.8& X & -40.74 $\pm$ 4.6 & -10.15 $\pm$ 1.3\\
\textbf{Most Visited} & 1.13 $\pm$ 0.1& 42.14 $\pm$ 4.3& X& 38.83 $\pm$ 3.5\\
\textbf{Blank UI}    &-36.47 $\pm$ 3.7& 11.98 $\pm$ 1.1& -36.86 $\pm$ 3.1& X \\ \bottomrule
\end{tabular}
\caption{Average ($\pm$ SD) percentage change in exploration between last session of old mode and first session of new mode for all participants. The $(i,j)^{th}$ entry represents transition from mode $i$ to mode $j$.}
\label{tab:speed}
\end{table}

While our experiment design is not meant to differentiate the possible influences of memory interference and motivational shifts based on changed presentations, some insight can be gleaned from the speed with which exploratory behavior changes with respect to UI changes. The part set cuing explanation for behavior change involves no learning or adaptation of preferences, whereas any sort of motivational effect on preferences would likely require learning over a somewhat extended time-scale. Table \ref{tab:speed} documents the average percentage change across these mode transitions, and shows that transitions from the default mode to either the least visited or blank UI mode caused about 80\% of the eventual change in exploration propensity to be manifest within the first session post-change. Such rapid transitions between exploratory behaviors contraindicates a learning-based explanation for the phenomenon, and support priming or interference based explanations. Figure~\ref{fig:temporal} shows the pattern of temporal variation in exploration of one individual over the course of the experiment.


\subsection{Followup analysis: weekday-weekend and time-of-day}

Forty two of our participants continued using our add-on for an extended period beyond the 60 day experiment window. We conducted some follow-up analyses on additional data we obtained from them to obtain a more complete picture of  comparative behavior in Blank UI mode versus the Default mode. These participants were subjected to the default mode UI for three weeks continuously and then to the Blank UI mode for seven weeks continuously. 

\begin{figure*}[thbp]
 \begin{subfigure}[b]{0.32\textwidth}
              \centering
    \includegraphics[width=\columnwidth]{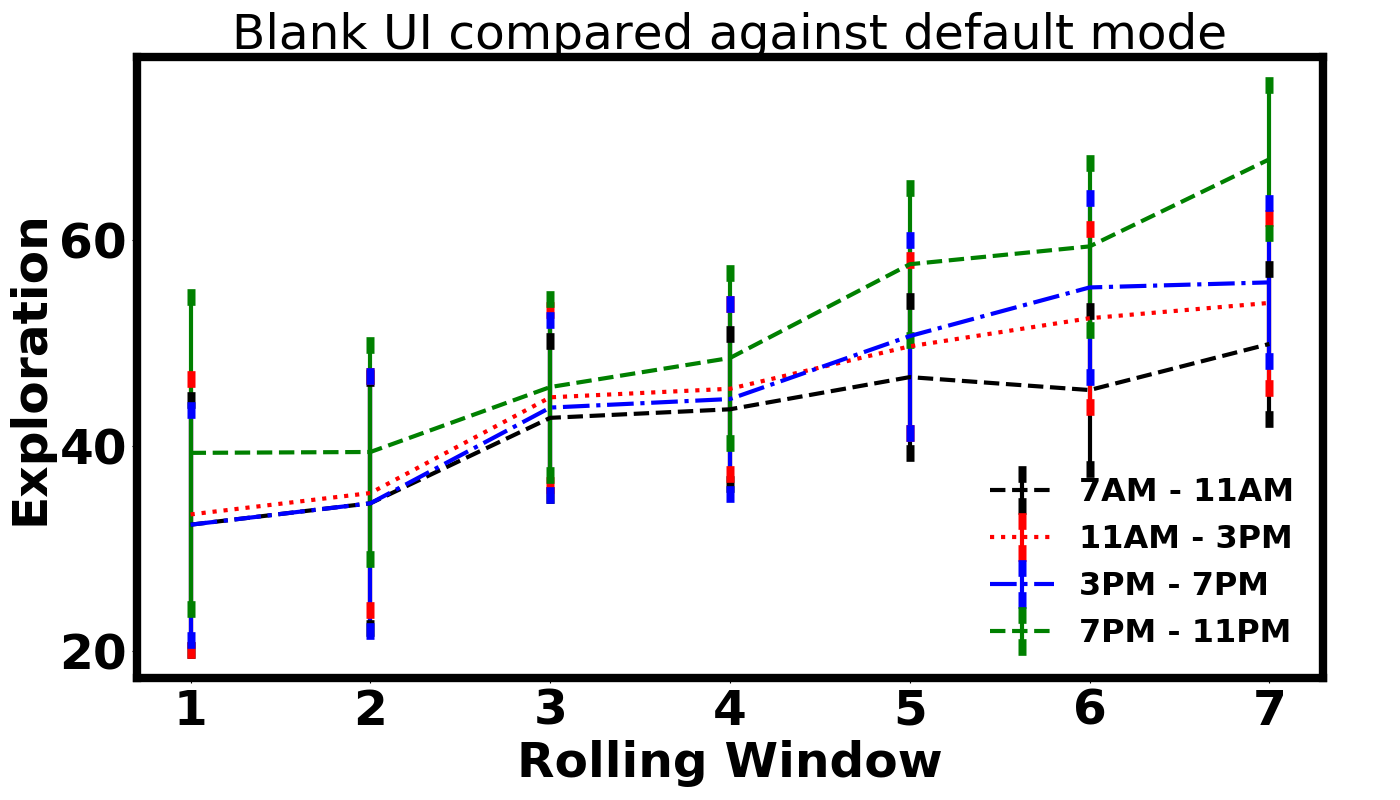}
    \caption{Exploration with time of the day}
    \label{fig:timeofday}
        \end{subfigure}%
        \begin{subfigure}[b]{0.32\textwidth}
              \centering
    \includegraphics[width=\columnwidth]{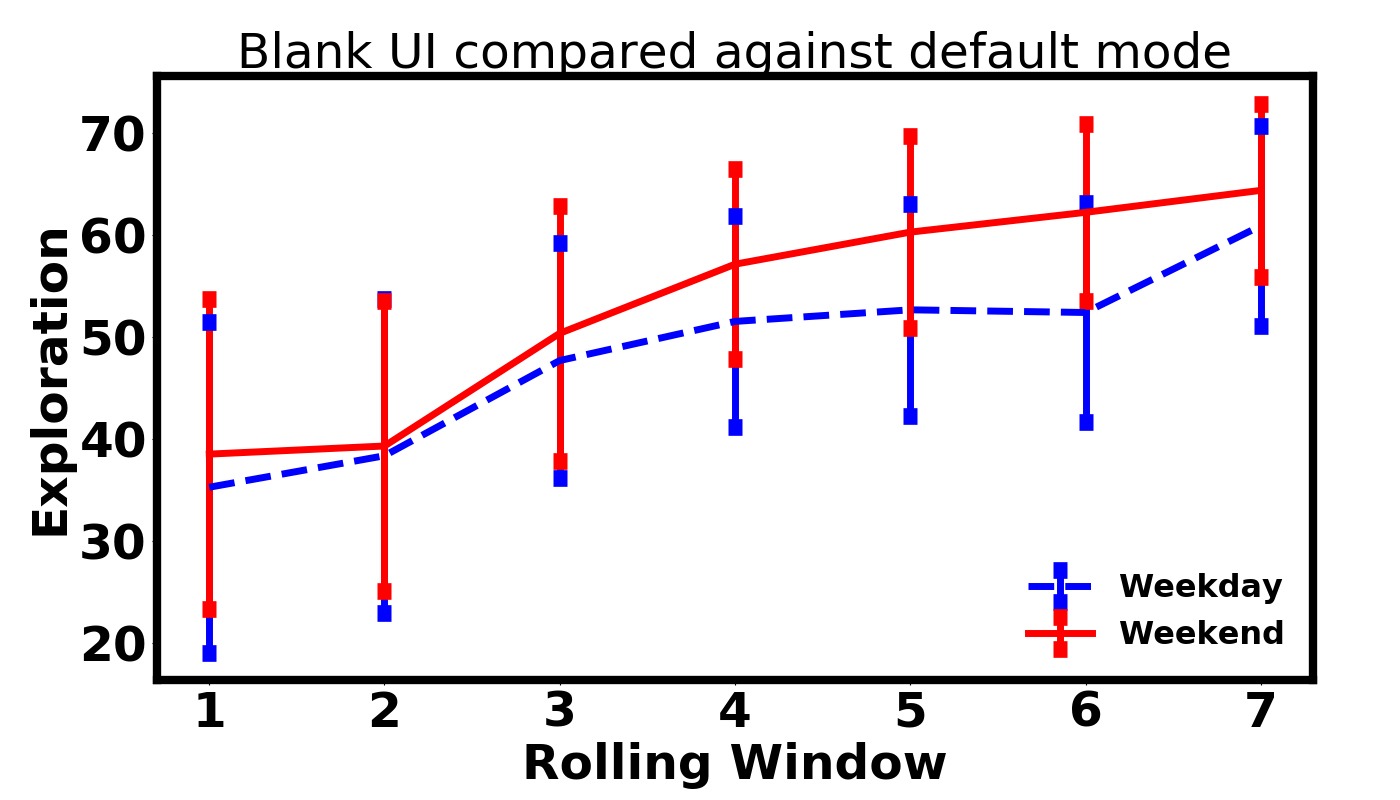}
    \caption{ Mean of exploration}
    \label{fig:weekmean}
        \end{subfigure}%
        \begin{subfigure}[b]{0.32\textwidth}
              \centering
    \includegraphics[width=\columnwidth]{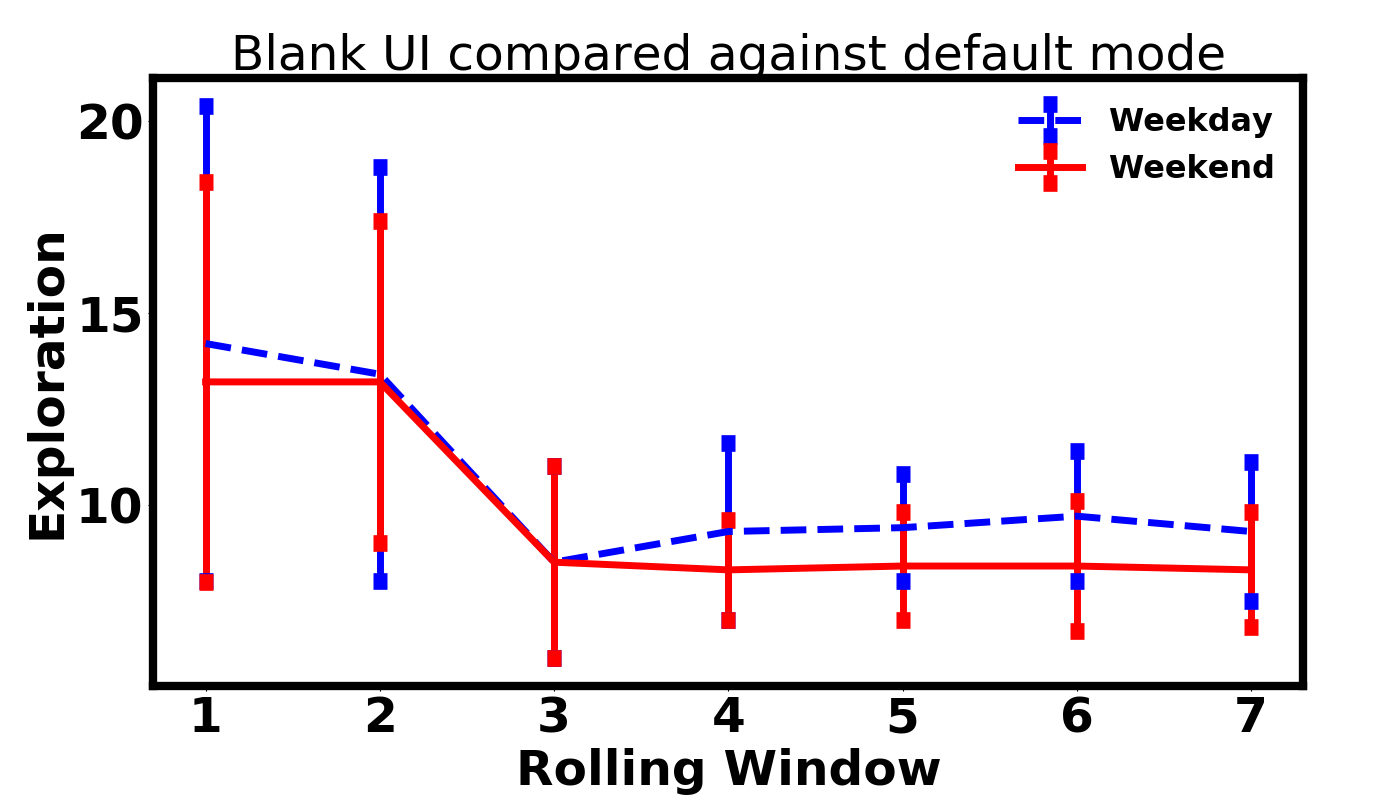}
    \caption{SD of exploration}
    \label{fig:weekvar}
        \end{subfigure}%

        \caption{Plot of \% increase in exploration for Blank UI mode over Default mode averaged over 40 participants. (a) Variation of Mean of exploration  over different time chunks of the day. (b) Variation of Mean exploration over weekdays/weekends (c) Variation of SD of exploration over weekdays/weekends 
        }\label{fig:additionalanalysis}
\end{figure*}

Using the data for default mode, we compute the base rate of \textit{exploration} at different blocks of time in the day per user. Now, we plot the percent change in \textit{exploration} in the Blank UI mode over default mode for each time block averaged over a rolling window of 7 days for the total duration of 7 weeks. Figure~\ref{fig:timeofday} shows this analysis averaged over all participants. The error bars represent Standard Deviation upon averaging the means over all participants. 

We observe that the percentage-wise increase in \textit{exploration} for the Blank UI mode over the Default mode increases with time for all time blocks. This is indicative of the conditioning due to tweaking of the new tab page UI from default mode to blank UI mode taking effect. This is an interesting observation, and contrasts with the rapid onset of change seen above in Fig. ~\ref{fig:temporal}. Combined, these observations appear to support a partial psychological role for both priming and conditioning in driving exploration suppression for web users, especially ones accustomed to using the default interface for long periods of time.  
We similarly plot the percent change in \textit{exploration} in the Blank UI mode over default mode for weekdays and weekends averaged over a rolling window of 5 days and 2 days respectively for the total duration of 7 weeks.  Figure~\ref{fig:weekmean} shows this analysis averaged over all participants. The error bars represent Standard Error of the Mean upon averaging the means per week over all participants. Figure~\ref{fig:weekvar} shows another picture of the "Weekday/Weekend" analysis averaged over all participants. For each participant in each week, the exploration per day is computed. The SD of exploration is computed for a 5 day block (weekday) and a 2 day block (weekend).  This plot shows how this SD varies over time averaged across all individuals. The error bars represent SD of the SD upon averaging across all participants. 

From Figures~\ref{fig:weekmean} and~\ref{fig:weekvar} we observe that the absolute value of increase in \textit{exploration} for the Blank UI mode over the Default mode increases with time. This is similar to the observation made in Figure~\ref{fig:timeofday}.

\vspace{0.1cm}
\section{Quantifying the consequences of exploration suppression}

\begin{center}
\begin{figure*}[htbp]
\includegraphics[width=0.99\textwidth]{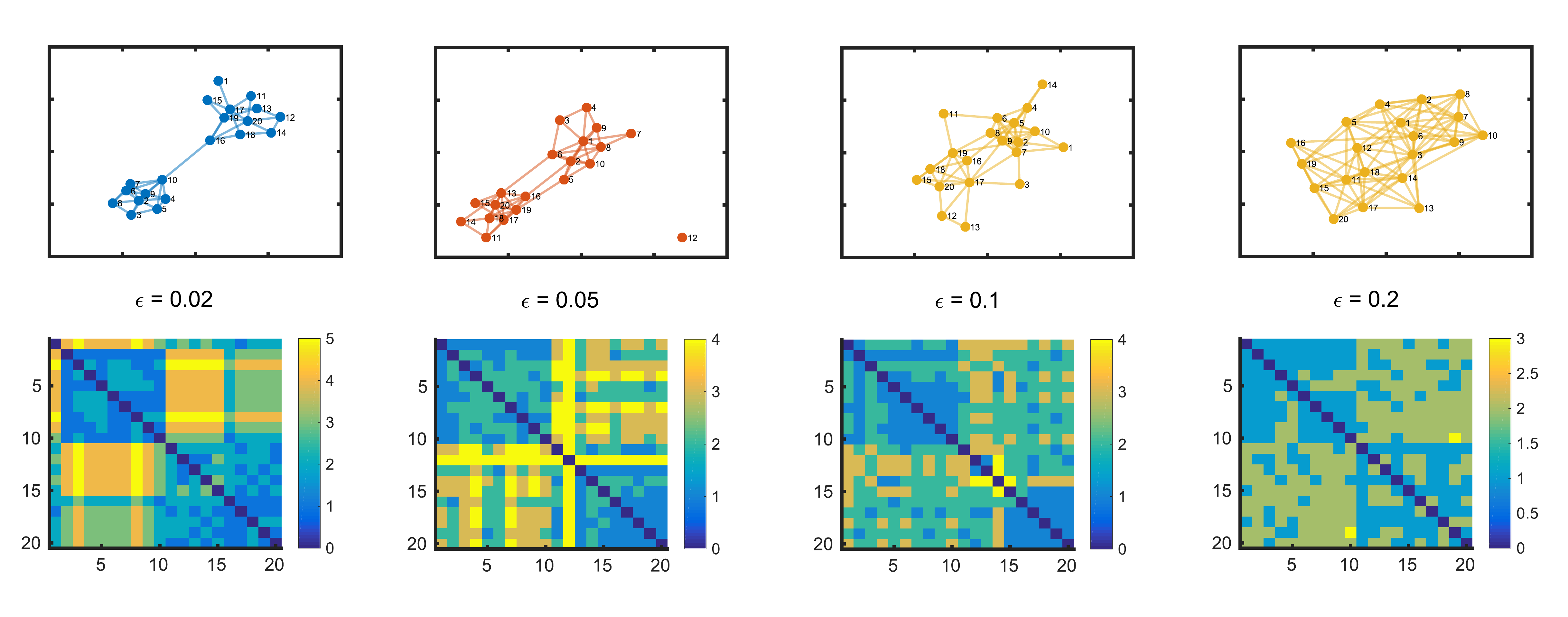}
\caption{ (Top row) Sample graphs with different $\epsilon$ values (Bottom row) Corresponding shortest paths between all node pairs in these graphs. The key point to note is that the connectivity of the graphs rises super-linearly as a function of $\epsilon$.}
\label{fig:graphs}
\end{figure*}
\end{center}

Our experiment shows that users' behavior adapts to manipulations of the browser new tab display along lines predicted by classic theories of memory interference, and that this adaptation is {\em statistically} significant both across and within individuals. We have not, however, established that this adaptation is also {\em practically} significant. {\em Prima facie}, the changes in behavior are only apparent at the margins, not in the large aggregate of browsing patterns in our subjects, which comprises overwhemingly of habitual access to highly visited websites. Viewed from this aggregate perspective, the deviations we have managed to introduce are of the order of about $0.5\%$ of all page visits. Why should this miniscule change in behavior be consequential? This is the question we have sought to answer using the {\em in silico} experiments we report below.

Our primary hypothesis in this phase of the project was that minor changes in exploration propensity would affect the facility with which users can access crucial {\em} transition nodes in their information network that connect them to fresh sources of information that they do not habitually encounter during routine browsing. This hypothesis was inspired by related observations about the role of social preference feedback in the polarization of internet communities - a phenomenon recently popularized by the term `filter bubble'~\cite{pariser2011filter}. 

In classic filter bubble settings, the source of feedback and recommendations is external, but tuned to personal preferences. It is this personalization that, perversely, reduces the diversity of information sources that the user acquires their social media-mediated knowledge from~\cite{pariser2011filter}. In our case, unlike in conventional filter bubble settings, there are no external sources of information and/or feedback. The browser UI is simply passively reflecting the habitual behavior of the user, modulo some modelling of the inter-temporal preferences for various forms of content of users.

Nonetheless, we think the filtering effect of reinforcement of existing preferences works by the exact same mechanism in our case, suppressing the user's propensity to sample content from diverse sources. Since the user's response to their own previous history is much more observable than social influences seen in social media recommender systems, we can try to quantify the extent to which suppression effects of the size seen in our empirical data might influence the diversity of information sourcing for model agents.

To do this, we have modelled website browsing behavior for an individual agent as a random walk on a graph. Within the graph for each subject, each node represents a distinct website. With this abstract representation in place, the next step is to realize that the structure of the graph itself must reflect the Zipfian nature of webpage-website counts - users visits a small number of sites a large number of times, and a large number of sites a small number of times~\cite{huberman1998strong}. 

To capture these particular dynamics while maintaining analytic simplicity, we specialized the network stucture we consider to be the family of all graphs with two pseudo-cliques wherein the cross-links between these pseudo-cliques are controlled by an $\epsilon$ parameter such that with a probability of $\epsilon$, a link gets set up between two cross-pseudoclique nodes\footnote{A pseudo-clique is defined as the graph structure obtained by removing some edges from a clique. We generate pseudo-cliques by first assuming an exponential degree distribution within a pseudo-clique of nodes, then assigning edges to nodes randomly in decreasing order of degree, permitting $\pm 1$ changes in the original degree assigned to each node to handle conflicts in edge assignment. In our experiments, we used pseudo-cliques of size N=10 with highest degree 4.}.  When we simulate a random walk on any one such graph, the reasons for this construction become clear. Nodes within pseudo-cliques have several short paths connecting them, and so are likelier to be visited by a random walker known to be situated at another node within the clique. Nodes in different pseudo-cliques will have longer shortest paths on average, so depending on the distribution of length of random walks for the agent, visits across pseudo-cliques will be rarer. Thus, we induce differences in accessibility of different nodes in the graph by assuming its structure to take this particular shape. 

Unlike more intricate agent-based models of browsing behavior, such as in~\cite{huberman1998strong}, which try to capture both within and across website browsing behavior, the model we use above makes predictions only for the quantity we are interested in - the across website browsing behavior of the user. Dwell times and page depths within websites are not modelled in our approach. We offload as much of the preference information about browsing behavior as possible onto the graph structure, resulting in the simplest possible observer model for the user - a random walk on the graph - that retrieves the Zipfian cross-site browsing behavior we are interested in simulating. The key difference between our simplistic model and the actual model of the web in terms of structure is that there are multiple pseudo-cliques in the web corresponding to different localizations of information~\cite{barbasi}. The results we demonstrate for two pseudo-cliques can be shown for any pairs of pseudo-cliques without loss of generality, thus generalizing to more complicated web models as in~\cite{huberman1998strong}.

The core of our succeeding analysis is identifying the slope of the relationship between the $\epsilon$ parameter, which controls the number of cross-links between the two pseudo-cliques, and the average first hitting time between nodes in different pseudo-cliques. Such an analysis does not require that the graphs contain pseudo-cliques of only equal size, or that they contain only two pseudo-cliques. We consider the simplest subset that instantiates the distance asymmetry we require. As a consequence of this distance asymmetry, we are able to map user behavior to network structure. Movements within a clique correspond to habitual behavior while a leap from one clique to another corresponds to exploration. Embedding the usually habitual, occasionally exploratory behavior of users in this way in our graph model, we can say formally that there is on average exploratory behavior $\epsilon$ of the time and habitual behavior $1-\epsilon$ of the time, assuming random initial node initialization in our specific graph structure. The same general principle is expected to hold for a much more general family of graphs - graphs with $k$ pseudo-cliques, interconnected by cross-links generated via a stochastic process mediated by the parameter matrix $\epsilon_{ij}, i,j\in \{1, \cdots k\}.$

In this setup we are interested, fundamentally, in quantitatively characterizing the relationship between changes in $\epsilon$ and changes in the average first hitting time between cross-pseudoclique nodes. The first hitting time between two nodes is simply the expected number of moves a random walker originating at node $i$ will take to first reach node $j$ across multiple random walks originating at $i$. Since the UI manipulation affects $\epsilon$ by a large relative amount but a small absolute amount, we are interested in measuring the extent to which this small parametric change affects the visibility of cross-pseudoclique nodes in the graph, which corresponds in our browsing user model to the average first hitting time. Users will access websites that lie within their radius of experience, as measured by hitting time. First hitting times larger than the typical radius of experience - measured in terms of walk length - will correspond to websites typically inaccessible to the user.  

Although theoretical properties of first hitting times on graphs have been investigated for random graphs~\cite{sood2004first,graphs,graphs2}, no direct results are specifically relevant for our analysis. Known theory suggests, on average, that adding extra nodes or edge to a graph increases the first hitting time of any two nodes in the graph but such analyses are true only on average across random graphs~\cite{aldous1989introduction}. We, on the other hand, are interested in graphs with a very specific direction of variation in structure, changing from barbell style graphs for low values of $\epsilon$ into well-connected graphs as $\epsilon$ rises. Figure~\ref{fig:graphs} shows various graphs generated corresponding to different $\epsilon$ values. The important thing to note is the rapidity with which adding cross-links changes sample graphs drawn generatively using different values of $\epsilon$ from barbell style graphs (known to have worst case shortest path distances) to well-connected graphs, with much smaller shortest path distances. Even though the graphs with higher $\epsilon$ have more edges, we intuitively expect that the non-random manner in which the new edges are introduced by our graph generative model counteracts the mere fact that they are being added, and reduces the first hitting time.
\begin{center}
\begin{figure}[htbp]
\includegraphics[width=0.95\textwidth]{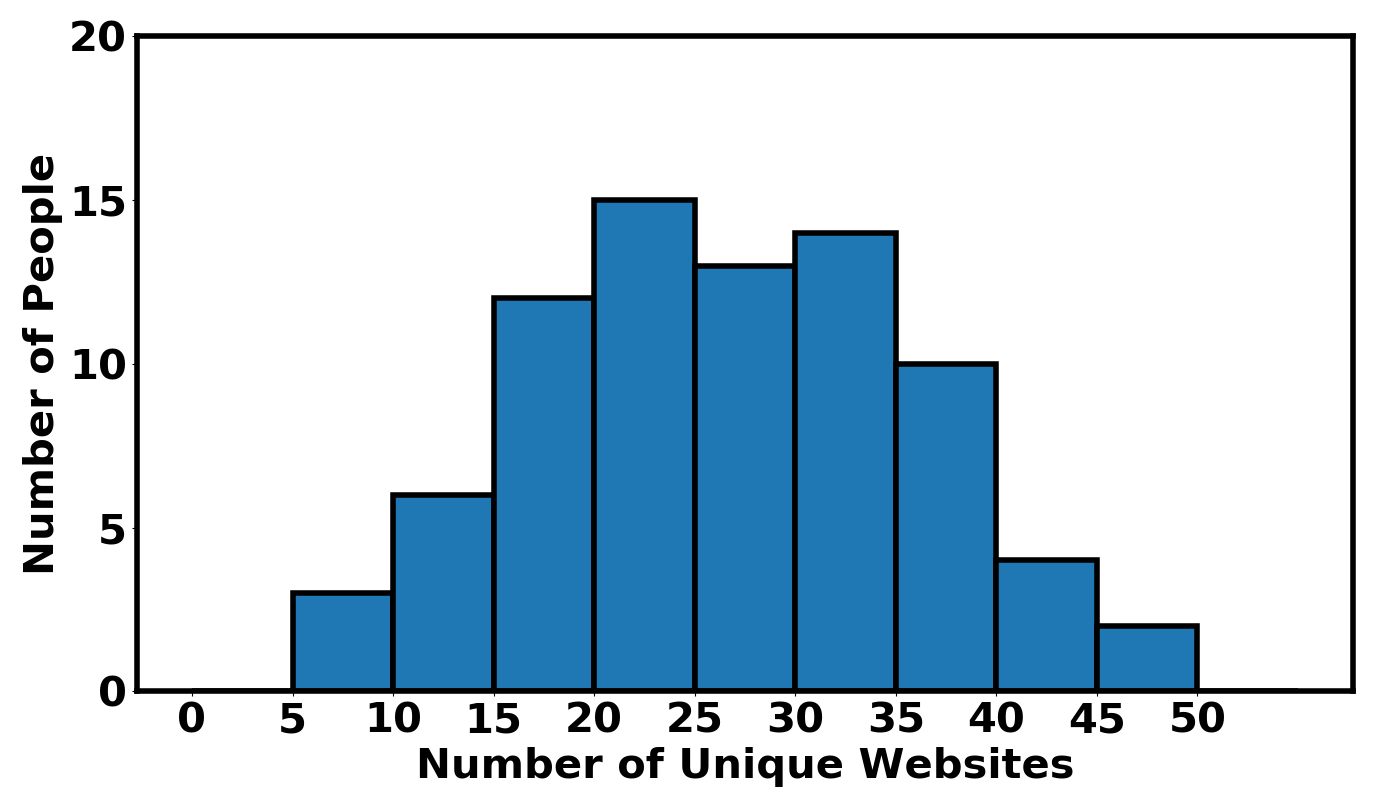}
\caption{Histogram showing the visit count of unique sites for the participants}
\label{fig:getn}
\end{figure}
\end{center}
\vspace*{-0.6cm}
\hspace{0.4cm}The generative model for our graph has two free parameters, the size of the graph $N$ and the cross-linking parameter $\epsilon$. We fit this generative model to our data by fixing the size of the graph and the value of the $\epsilon$ parameter using the typical size of our users' unique website repertoire and their measured exploration/habit ratios in our dataset respectively. The modal value of $N$ (the number of unique sites visited by a user in a month) as inferred from the histogram in Figure~\ref{fig:getn} is $25$. To approximate this, each of our simulated graphs used $20$ nodes divided into two pseudo-cliques of equal sizes. We generated graphs using $\epsilon$ values varied between 0 and 0.2 in steps of 0.005. 

\begin{center}
\begin{figure}[htbp]
\includegraphics[width=0.95\textwidth]{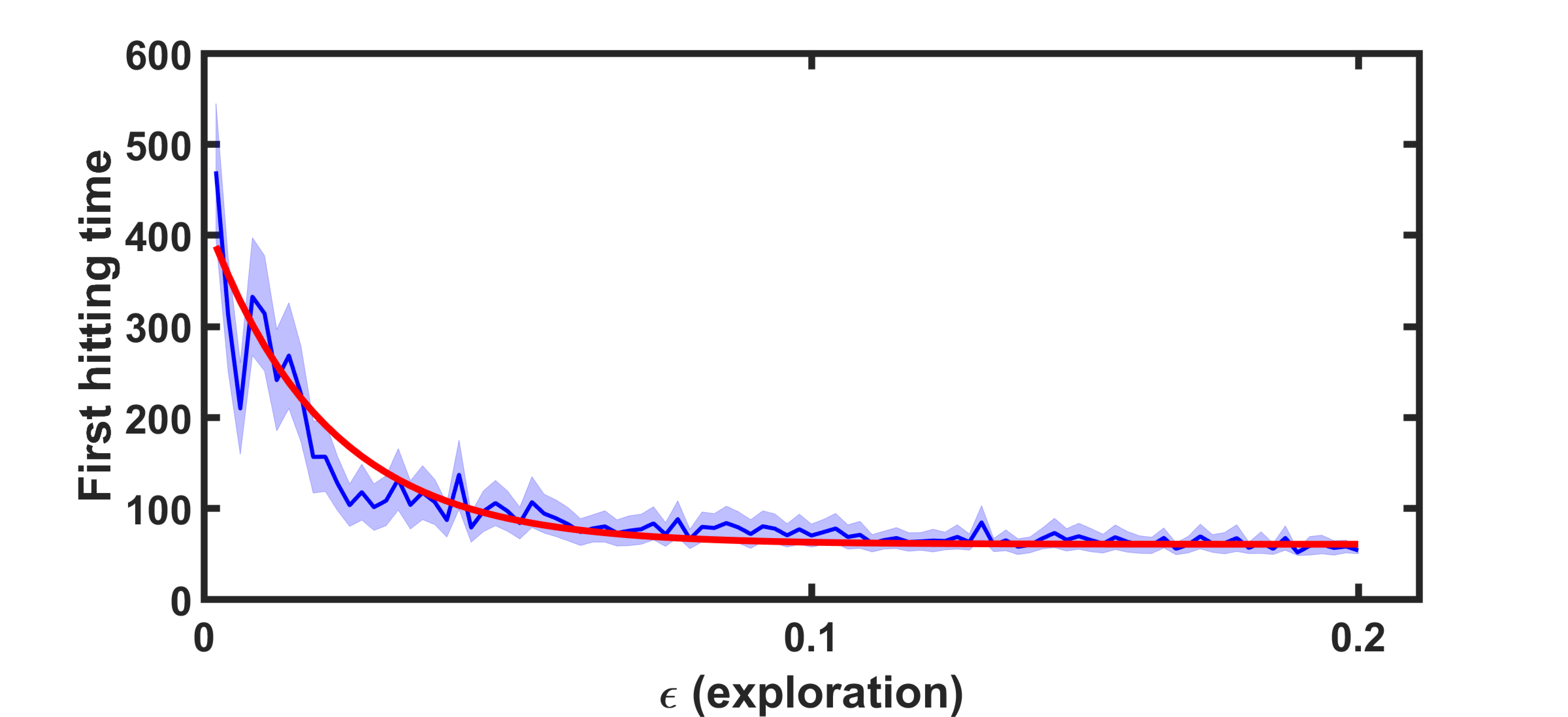}
\caption{Plot of average and $95$\% confidence bound of first hitting time for all cross-pseudoclique nodes empirically measured across 100 simulated random walks in each of 100 graphs generated for each value of $\epsilon$. The best fit exponential for the data is also shown.}
\label{fig:hitting}
\end{figure}
\end{center}

Once an $(N,\epsilon)$ graph for a particular value of epsilon is ready, we empirically measured the hitting time of every cross-pseudoclique node pair by simulating random walks between all cross-pseudoclique node pairs. We calculated the average first hitting time for 100 such random walks across all cross-pseudoclique node pairs for each of 100 $(N,\epsilon)$ generated graphs for each $\epsilon$ value. Figure \ref{fig:hitting} shows the result of this simulation, plotting average first hitting times obtained from this experiment for different values of $\epsilon$.

We found empirically that the range of $\epsilon$, measured as the ratio of exploration to habit pages for each subject, is $0.005-0.07$ across our subjects. In this range, as illustrated in Figure \ref{fig:hitting} we observe an approximately exponential decline in hitting time with increase in $\epsilon$. This is because, as anticipated, the effect of the `cliqueish' structure of the graph dominates over increase in number of edges with $\epsilon$. Gradually, as $\epsilon$ increases, the effect of increase in number of edges between pseudo-cliques starts dominating and hence the average hitting time asymptotes beyond $\epsilon\approx 0.1$, and will likely increase for still higher values. The key finding, also as anticipated, is that within the range of parametric values seen in our experiment, even small changes in $\epsilon$ cause large changes in the first hitting time. As we describe above, first hitting times represent website accessibility in our model, so the result obtained here implies that small changes in $\epsilon$ cause exponential declines in the accessibility of infrequently visited (cross-pseudoclique) websites for users.
\hspace{0.4cm}Finally, the relationship between a finite random walk and the probability that it will pass through any given node is straighforward to estimate. It is simply the probability that a random walk as long as $N$ might occur in the empirical distribution of first hitting times for that specific node. Hence, the p-value of a one-sided probability hypothesis test will give us exactly the probabilistic quantity of most interest to us - the probability with which a user with exploration propensity $\epsilon$ will access a diverse information source in any given browsing session of any particular estimated length. 

We identified individual sessions in our participants' browsing logs using a 30 minute interval to determine the boundaries of individual sessions (as mentioned in Section 2.3), and counting total page visits as one step in the random walk. Across all participants, the average session length came out to be around $20$ for our dataset. Next, we measured the kurtosis of the histogram of hitting times for each node pair in multiple random walks on a particular graph. We found that the kurtosis values themselves were well-represented by a normal distribution with mean 3.1 and SD 0.2. Hence, the first hitting time distribution for each node pair could be considered approximately Gaussian.

\begin{center}
\begin{figure}[htbp]
\includegraphics[width=0.95\textwidth]{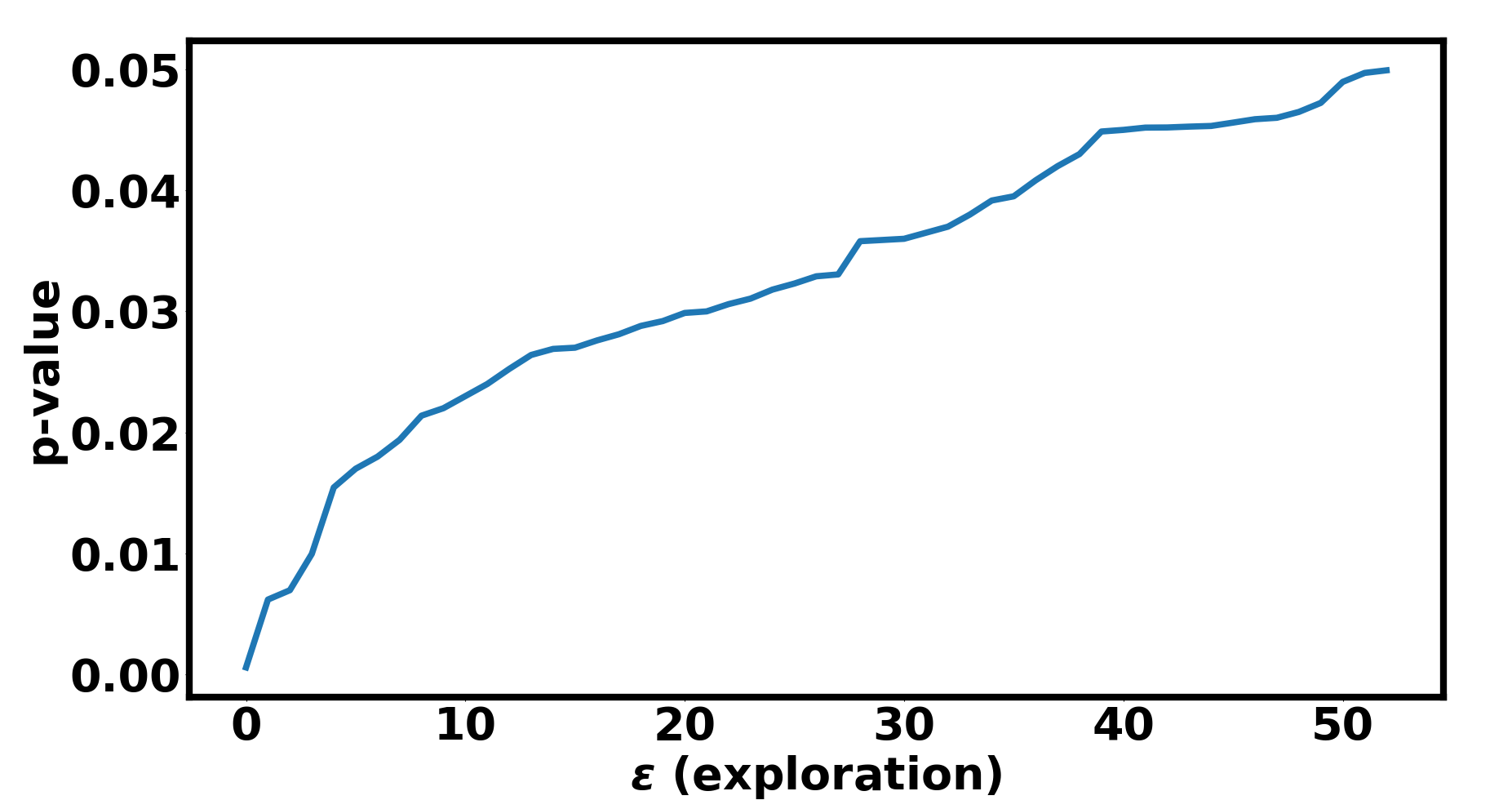}
\caption{Variation of p-value of the average session length with exploration. p-value is averaged across multiple cross-pseudoclique node-pairs and graphs.}
\label{fig:pval}
\end{figure}
\end{center}

So, we ran one-sided z-tests testing whether a session length of $k = 20$ or lower might occur naturally in the hitting time distribution of cross-clique node pairs. We ran this calculation for 100 randomly sampled cross-pseudoclique node pairs across 100 graphs generated for each value of $\epsilon$ and plotted the p-values obtained against $\epsilon$. This relationship is visualized in Figure \ref{fig:pval}, and is approximately linear. This observation quantifies concretely the importance of large relative changes in exploration propensity, even if exploration constitutes a small relative share of overall browsing behavior.  While the change in behavior is on the margin in terms of aggregate browsing behavior, it has non-marginal consequences on which information sources users will be able to access.

\section{Discussion}

Since the beginning of internet browsing, designers have always tried to design browsers in ways that reduce the user's typing effort. In early browsers, this was done using history auto-complete suggestions. More recently, both OS and browser designers have sought to reduce typing effort still further by adding apps to home screens and frequent (and recent) web pages to `new tab' pages. The inarguable logic of such design is that, because website visit counts are approximately power law distributed, being able to simply click on the most visited pages optimizes the number of typing responses needed over the course of users' use of the browser.

But website visits are a means to an end, not an end in themselves. Ultimately, they are expressions of our information preferences~\cite{pirolli1995information}. This paper makes the case that, by showing people the sites they visit most frequently over and over again in new tab displays, current practice in browser and OS UI design traps people into a solipsistic feedback loop, reinforcing their strongest preferences to the detriment of weaker ones that potentially offer more scope for diversity of experience and learning~\cite{schmidhuber2010formal, srivastava2011cognitive}. We quantified the size of this feedback effect at its source - natural browsing behavior - by collecting browsing history from subjects who consented to having the behavior of their new tab page manipulated remotely via a plug-in they installed in their browsers. With in silico experiments, we further demonstrated the impact of the change in browsing behavior on peoples' proclivity for acquiring information from diverse sources. 
\subsection{Related work}

Our work follows a rich vein of empirical research in human-computer interaction that is increasingly discovering subtle but powerful ways in which website interaction, access and even simple UI decisions can influence human decisions. For example, posting to Facebook has been shown to increase users' activity on the website immediately before and immediately after the posting, on the timescale of days~\cite{grinberg2016changes}. More substantively, time of day of Twitter activity has been shown, strikingly, to correlate strongly with the probability of users being clinically depressed~\cite{de2013predicting}. Even more strikingly, Epstein and colleagues have shown that presenting potential voters with search results for potential candidates in an actual national election in a UI with artificially manipulated search result rankings can shift the voting pool's vote fraction by up to 2\% in the experimenter's chosen direction,  a shift large enough to sway a reasonably close election~\cite{epstein2015search}.  There have also been recent works studying memory recall in simple tasks like clicking of pictures through different capture modalities~\cite{niforatos2017can}. We identify a large and consequential psychological effect: conventional UI design choices suppress peoples' propensity to access diverse information sources while accessing the web. This finding is not as specific as the ones documented previously in the literature, but also as a consequence, is likely affecting a lot more people at any given point in time by virtue of its generality. 

Our results also relate to research efforts ongoing to characterize and surmount the recommender systems (RS) {\em filter bubble}~\cite{filter1,filter2,filter3}. There is mixed evidence regarding the effects of recommender systems on users' consumption patterns. Whereas work focused on identifying changes in content {\em supply} as a function of personalization tools has found no effects~\cite{haim2018burst}, other work studying {\em consumption} patterns have reported both negative~\cite{flaxman2016filter, filter1} and positive effects on consumption diversity~\cite{nguyen2014exploring}. Our work finds unambiguous and large negative effects on information diversity driven by the use of recommendations. By our simplest measure of diversity - count of unique websites visited - users operating browsers with blank new tab pages visited on an average 15\% more unique websites over a two month period.  

One possible reason for previously reported null or positive results in the literature might be that the {\em active} use of recommender systems like Movielens involves users actively looking for information they don't already possess, an explicit cognitive task where they are aware that they must evaluate the incoming information quasi-critically~\cite{sinha2002role}. In contrast, constantly stopping by the new tab page during transitions between websites is a much subtler phenomenon; the recommendations here are {\em passive}, in the sense that the ostensible function of these bookmarks is to facilitate access to frequently visited pages, not to shift preferences. People can reason through, or actively ignore, redundant information while they are actively deciding what to do. It is harder for them to guard against the subtle impact of memory inhibition through visual presentation of web page icons time after time in the normal flow of their interaction with the browser~\cite{newell2014unconscious}. 

While it might seem superficially strange that such a small UI design component could potentially have such a large impact on browsing behavior, behavioral economics offers striking examples of similar subconscious nudges significantly affecting behavior for better and for worse~\cite{leonard2008richard}. Similarly, Thaler and Sunstein offer striking examples of how small and often insignificant things can influence (nudge) behavior~\cite{thaler-nudge}. As McLuhan foresaw, but many modern empirical analyses ignore, the peril in informative computer interfaces is not that they may feed us bad information against our will, it is that the {\em form} in which they present information to use will subtly change our expectations of what we want to look for. 
The classic ideal observer in web browsing  - Pirolli \& Card's \textit{information forager} - models humans as observers optimizing information acquisition under information-theoretic constraints~\cite{pirolliforaging,pirolli1995information,pirolli2003snif}. Our demonstration of large priming effects caused by as innocuous a source as the new tab page interface could help further nuance such models by incorporating the effect of psychological biases in the same way the influential cognitive biases literature has influenced the design of microeconomic models~\cite{tversky1974judgment}. 

Our description of the delicate interaction between mind and machine in this most humdrum of tasks - web browsing - should also stimulate further exploration of the negative externalities of personalization, recommender systems and other methods of preference influence common in HCI applications. At present, such influence is sometimes justified by discriminating between malevolent and benevolent deception in UI design, with the basic difference being that a benevolent deception is meant to benefit the user~\cite{adar2013benevolent}. Our study presents a clear example of the inadequacy of such ethical definitions - in our case, the developer, in principle, gains nothing by presenting frequently viewed web pages as clickable icons; the UI is designed the way it is to minimize the user's need to type. Nonetheless, this benevolent deception, conflating accessibility with preference via the subconscious impact of presentation on memory, ends up affecting user behavior substantially, as our results show.

\subsection{Limitations}

A natural limitation of this work is the size and representativeness of the population sample we used to establish the basic fact that presentation of commonly used websites' page icons on the new tab display interferes with the retrieval of infrequently viewed websites.  Nonetheless, the empirical results we describe in this paper are statistically significant by all conventional standards of measurements, and our sample is quite diverse with respect to gender, education, and occupational status. While one can never have too many participants in a study making claims about internet browsing behavior, the large effect sizes we find, consistently seen both across and within participants, suggest that our study was adequately powered to discern the specific hypothesis it investigated.  


It is also possible to question how well conclusions from this study, with a mean sample age of 28, may generalize to the experience of older adults. At the same time, it must be remembered that while our sample is not representative of the general population, it is certainly more representative of {\em heavy} internet users~\cite{assael2005demographic,internetusers}, who are the population most likely to be affected by the UI design decisions critiqued in this paper. We also note that the average age of our population is very close to the median age of the country this study was conducted in, suggesting that there is no significant age-bias in our recruitment. 

 Our results show that changes in the new tab page interface affect the diversity of users' browsing experience, and that this change is brought about specifically by interacting with the new tab page. But they are unable to shed light on the psychological phenomenon underlying this effect. Whereas we initiated this experiment inspired by the analogy between web browsing and memory retrieval, successful retrieval from memory requires both motivation and memory performance. Our experiment design does not differentiate these two variables. Thus, we cannot claim rigorously that it is memory impairment caused by competitive interference from displayed web page icons that is driving the difference in website browsing patterns across our manipulations. We can merely suggest that this is likely to be true, to the extent that motivation to access different websites does not change as a function of the UI display modes and offer as supporting evidence the fact that browsing behavior changed in the same direction and by about the same amount whether infrequent websites or no websites were shown on the new tab page. Identifying the mechanism by which the UI changes are transforming into behavioral changes is essential to identify useful solutions.  

\subsection{Conclusion}

In this paper, we contribute two basic results. One, we show empirically that subtle UI manipulations of the new tab page in common browsers can create large relative changes in users' propensity to explore sites outside the repertoire of sites they habitually visit. Two, we show using a simulation study informed by empirically measured parameters that large relative changes in exploration propensity manifest linearly in large absolute changes in the visibility of diverse information sources in a user's browsing experience. Together, these results present striking evidence that the current design practice of showing frequently visited webpages on new tab pages is suppressing the expression of web surfers' exploratory tendencies on the web. 

Considering how ubiquitous computers and the internet have become, this effect would span even mobile devices. In fact, since small screen sizes in hand held mobile phones or tablets are likely to bias users towards clicking the new tab page icons over typing in the address bar, the exploration suppression is likely to be exacerbated, as an admittedly small {\em post hoc} cohort analysis of our data shows. Further research is needed to quantify the extent of this difference.

Finally, we note that all common browsers offer users the ability to use a blank page as their new tab page, making our conclusions immediately actionable. Based on our results, people should decide for themselves whether the convenience of clicking through to commonly visited web pages is worth the potential curtailment of their curiosity. 

%
\bibliographystyle{acm}
\bibliography{sample}

\end{document}